\documentclass[10pt]{elsarticle}

\usepackage{amsthm,amsmath,amsfonts,amssymb,amscd,mathrsfs,comment}
\usepackage{txfonts}
\usepackage{supertabular,soul}
\usepackage{longtable}
\usepackage{multirow}
\usepackage{graphicx,color,geometry}
 \usepackage{multirow}
 \usepackage{graphics}

\usepackage[utf8]{inputenc}
\usepackage{graphicx}
\usepackage{subfig}
\usepackage{bbm}
\usepackage{hyperref}
\usepackage{yfonts}
\usepackage{eucal}
\usepackage{overpic}
\usepackage{enumitem}
\usepackage{dsfont}
\usepackage{soul}
\usepackage{natbib}
\usepackage{float}

\newcommand{\vertex}[1]{#1}
\newcommand{\edge}[2]{\{#1,#2\}}

\newcommand{\Pinterval}[2]{%
    \mathchoice%
    {\left[#1, #2 \right)}
    {[#1, #2)}
    {[#1, #2)}
    {[#1, #2)}
}%
\newcommand{\bigsimplex}[1]{[\begin{matrix} #1 \end{matrix}]}
\newcommand{\tetrahedron}[4]{\bigsimplex{#1 & #2 & #3 & #4}}
\newcommand{\triad}[3]{\bigsimplex{#1 & #2 & #3}}

\newcommand{\PHrep}{representative}
\newcommand{\PHreps}{representatives}
\newcommand{\horcycle}{union cycle}
\newcommand{\Horcycle}{Union Cycle}

\newcommand{\NN}{\mathbb N}
\newcommand{\ZZ}{\mathbb Z}
\newcommand{\R}{\mathbb R}

\newtheorem{definition}{Definition}
\newtheorem{thm}{Theorem}[section]

\newtheorem{remark}[thm]{Remark}

\newtheorem{example}[thm]{Example}

\theoremstyle{remark}

\let\oldmarginpar\marginpar
\renewcommand\marginpar[1]{\oldmarginpar[\raggedleft\footnotesize #1]%
{\raggedright\footnotesize #1}}

\makeatletter 
\def\ps@pprintTitle{
  \let\@oddhead\@empty
  \let\@evenhead\@empty
  \let\@oddfoot\@empty
  \let\@evenfoot\@oddfoot
}
\makeatother

\begin{document}

\begin{frontmatter}

\date{\today}

\title{The persistent homology of genealogical networks}
\author[zach]{Zachary M. Boyd\corref{cor1}}
\cortext[cor1]{Corresponding Author}
\author[nick]{Nick Callor}
\author[taylor]{Taylor Gledhill}
\author[abby]{Abigail Jenkins}
\author[robbie]{Robert Snellman}
\author[ben]{Benjamin Webb}
\author[rae]{Raelynn Wonnacott}
\address[zach]{Department of Mathematics, Brigham Young University, Provo, UT 84602, USA, zach\_boyd@byu.edu}
\address[nick]{Department of Mathematics, Brigham Young University, Provo, UT 84602, USA, n.b.callor@gmail.com}
\address[taylor]{Department of Mathematics, Brigham Young University, Provo, UT 84602, USA,  gledhilltaylor2@gmail.com}
\address[abby]{Department of Mathematics, Brigham Young University, Provo, UT 84602, USA, jenkins.abby@gmail.com}
\address[robbie]{Department of Mathematics, Brigham Young University, Provo, UT 84602, USA, snellman@mathematics.byu.edu}
\address[ben]{Department of Mathematics, Brigham Young University, Provo, UT 84602, USA, bwebb@mathematics.byu.edu}
\address[rae]{Department of Mathematics, Brigham Young University, Provo, UT 84602, USA, raelynnwo@gmail.com}


\begin{abstract}
Genealogical networks (i.e. family trees) are of growing interest, with the largest known data sets now including well over one billion individuals. Interest in family history also supports an $8.5$ billion dollar industry whose size is projected to double within 7 years [FutureWise report HC-1137]. Yet little mathematical attention has been paid to the complex network properties of genealogical networks, especially at large scales.

The structure of genealogical networks is of particular interest due to the practice of forming unions, e.g. marriages, that are typically well outside one's immediate family. In most other networks, including other social networks, no equivalent restriction exists on the distance at which relationships form. To study the effect this has on genealogical networks we use persistent homology to identify and compare the structure of 101 genealogical and 31 other social networks. Specifically, we introduce the notion of a network's persistence curve, which encodes the network's set of persistence intervals. We find that the persistence curves of genealogical networks have a distinct structure when compared to other social networks. This difference in structure also extends to subnetworks of genealogical and social networks suggesting that, even with incomplete data, persistent homology can be used to meaningfully analyze genealogical networks. Here we also describe how concepts from genealogical networks, such as common ancestor cycles, are represented using persistent homology. We expect that persistent homology tools will become increasingly important in genealogical exploration as popular interest in ancestry research continues to expand.
\end{abstract}


\end{frontmatter}

\section{Introduction}
The study of genealogical networks, that is networks relating parents with children and spouses with each other through successive generations is of rapidly growing interest, both because of genealogy's popular appeal and its applications in genetics~\cite{longevity}, sociology~\cite{bicomponents}, population sciences~\cite{chang_2004}, and economics~\cite{inequality}. Growing data availability of rich, temporally resolved data is also driving interest in genealogy. For example, FamilySearch has constructed a human family tree with over $1.40$ billion individuals, based on $2.21$ billion sources, including $4.78$ billion images (https://www.familysearch.org/en/newsroom/company-facts). Popularization of DNA testing services and increasing availability of audio sources, geographic tags, occupation metadata, and migration records combine to make genealogical networks some of the largest, most richly featured, geospatially embedded temporal networks in existence. Examples of relevant academic studies include methods for automatically constructing networks from documents~\cite{finnish,population_reconstruction}, analyzing marriage patterns~\cite{inequality}, structured population modeling, branching processes~\cite{structured_populations}, and biconnected components~\cite{bicomponents, group_theory}. Of particular interest to us are works that study distance to recent common ancestors, both theoretically and via simulation (e.g.~\cite{chang_1999,chang_2004}). A growing body of literature also uses genealogical networks for genetic inference, as in~\cite{longevity}.

Related to these genealogical endeavors, a major goal of network science is to describe the structure of such real-world networks. In this paper, we consider persistent homology as a tool to both analyze and explore the structure of genealogical networks. Persistent homology, roughly speaking, is a method of representing voids or gaps in the structure of a network, that distinguishes how significant these voids are to the overall network structure. Persistent homology can be used to compare these voids across two networks without requiring a correspondence between the individual vertices or edges, or even requiring the networks to be the same size. The basic idea involves ``filling in'' the network with simplices (points, edges, triangles, tetrahedra, etc.) and keeping track of how the network changes as we do so (see Section~\ref{sec:3} for details).


Some similar applications of persistent homology in the study of networks include \cite{Carstens}, \cite{kannan2019}, \cite{D09}, \cite{PSDV13}. The collaboration networks studied in \cite{Carstens} are similar to the social networks that we use for comparison in this paper, though our focus is primarily on distinguishing these from genealogical networks. Both \cite{kannan2019} and \cite{D09} apply persistent homology techniques to general randomized networks of various forms. It is also possible to vary the technique for generating a topological object from a network, as in \cite{PSDV13} where three methods are compared. We also recommend \cite{AAF19} and \cite{Otter17} as good overviews of the general methods of applying persistent homology.

For this paper, our method of constructing a topological representative for each network follows the same general pattern as the work cited above. However, we also acknowledge the wide variety of alternatives for encoding such information. \cite{CGOS13} and \cite{VDS18} encode their information as point-clouds rather than graphs. A higher-dimensional version of persistent homology is presented in \cite{BL20}, which may permit the inclusion of time-varying networks. Finally, the formulation in \cite{ABB20} may allow for better analysis of corrupted or too-large datasets.

We also wish to bring attention to four particular applications that demonstrate the versatility of persistent homology. In each of these applications, persistent homology has been used to identify structural voids in data and then to associate these voids to recognizable features in the underlying networks. It is the latter use that we wish to emphasize. Robins et al.~\cite{Robins2016} have shown that voids found using persistent homology correspond to percolating spheres in a porous material. In \cite{Lee2012}, structural voids arise when several groups of neurons are strongly connected sequentially, but out-of-sequence pairs are only weakly connected. In these neurological networks, persistent homology provides a way to identify and classify these different sequences as well as quantify the strength of these connections. The application in \cite{Duman} provides a method for extending traditional genetic analysis tools to a parameterized family of datasets by constructing an appropriate topological object. Lastly, \cite{BBDF16} shows that structural voids or gaps can also represent much more abstract concepts. In this case persistent voids are shown to correspond to the atonality in music compositions.

Intuitively, the voids or gaps in genealogical networks should be quite different when compared with other networks, such as social networks, since unions\footnote{In order to be inclusive of various relevant relationships in this paper, we use the word ``union'' to describe not only legal marriages and common law marriages but also some others, including any relationship that produced children.} (such as marriages) in genealogical networks typically form at specific distances, rather than through other mechanisms e.g. triadic closure. That is, distances between individuals who form unions are typically not too small or too large (see Section \ref{sec:2}). In contrast, in other social networks, new connections can form at any distance but are often quite small \cite{Sintos2014}. This difference in network growth between genealogical and other social networks causes differences in network topology that are reflected in the network's persistent homology. Thus persistent homology is a useful descriptive tool for exploring and modeling the structure of genealogical networks.

Here, we propose a new method for representing persistent homology, which we call a persistence curve (see Section \ref{sec:4}). The persistence curves of many genealogical networks are very similar to each other, and importantly the persistence curves of subsets of genealogical networks, that is, \textit{sampled} genealogical networks, are also similar to the persistence curves of unsampled genealogical networks (see Section \ref{sec:results}).

To give our study of genealogical networks context we also study the persistent homology of social networks. We find that the same result holds for the social networks we consider, in that the persistence curves of social networks show a common pattern and the persistence curves for social and sampled social networks are similar (see Section \ref{sec:results}). We confirm our analysis using another tool for comparing persistent homologies, the bottleneck distance, which is also capable of detecting and differentiating the distinct homology patterns between genealogical and other social networks.

In summary, we make the following contributions:
\begin{itemize}
    \item Introduce the notion of a persistence curve and introduce the use of this together with the bottleneck distance as a tool for the analysis of general networks.
    \item Report the distinct persistent homology structure of genealogical networks using both persistence curves and the bottleneck distance.
    \item Link this structure to genealogically relevant concepts.
    \item Similarly, report the distinct persistence homology structure of social networks and compare this to the structure of genealogical networks.
    \item Report evidence that persistent homology methods work well even in the presence of incomplete data. This is particularly relevant given that genealogical data is often, if not necessarily, incomplete.
\end{itemize}
Throughout the paper, examples from family networks are contrasted with other social networks to highlight the unique features of genealogical networks from a persistent homology point of view.

The paper is organized as follows. In Section \ref{sec:2} we describe both genealogical and social networks. In Section \ref{sec:3} we define the persistent homology of a network and introduce the notion of persistence curves. In Section \ref{sec:4} we define the bottleneck distance and show how both this distance and persistence curves can be used to compare networks. In Section \ref{sec:5} we describe the genealogical and social data sets we use in our study and give our experimental results in Section \ref{sec:results}. Section \ref{sec:results} also includes a discussion of how certain structural features of social and genealogical networks are represented using persistent homology. In Section \ref{sec:conc} we summarize our results and conclude with a discussion regarding the use of persistent homology as a tool for analyzing general network structure and recovering network features. Throughout we give examples of each of the concepts we introduce.

\section{Background: Genealogical and Social Networks}\label{sec:2}
We represent genealogical networks with a graph $G=(V,E)$, where $V=\{1,2,\dots,n\}$ are the individuals within the network, and $E$ are the (genealogical) relationships. These relationships consist of both \textit{parent-child} edges and spouse (or more generally \textit{union}) edges. For the sake of simplicity, these edges are considered to be undirected.
\begin{figure}[ht]
\begin{center}
    \begin{tabular}{c}
    \begin{overpic}[scale=0.3]{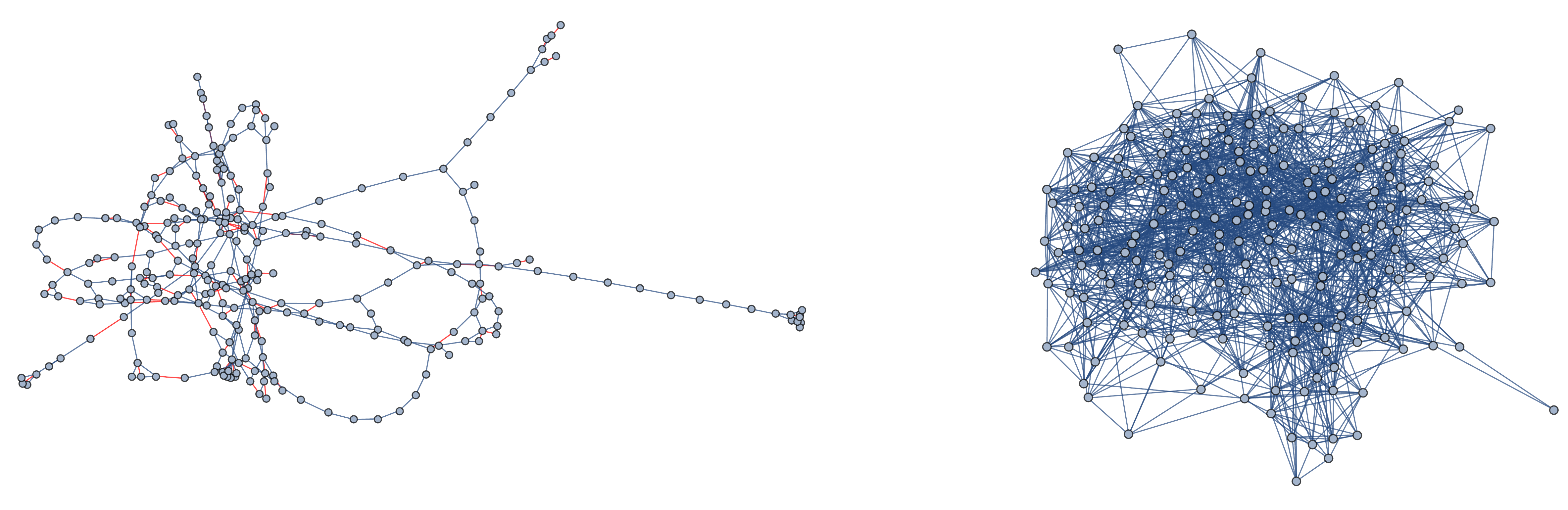}
    \put(10,-3){Tikopia Genealogical Network}
    \put(65,-3){Residence Hall Social Network}
    \end{overpic}
\end{tabular}
\vspace{0.25cm}
\caption{Left: The largest connected component of the Tikopia genealogical network consisting of 288 individuals from the island of Tikopia in Polynesia from the year 1930 to 2008, is shown \cite{kin21}. Parent-child edges are shown in blue and union edges are shown in red. Right: The largest connected component of the Residence Hall social network consisting of 217 individuals and their friendships from the Australian National University campus is shown \cite{rh}.}\label{fig:1}
\end{center}
\end{figure}
We note that the structure of a genealogical network is often thought of as being ``tree-like'', since genealogical networks are often constructed from an individual, their parents, their grandparents, and so on, ignoring union edges. The result is a \textit{tree}, i.e. a connected acyclic graph, if we create only a few generations of the family. However, full genealogical networks are not trees due to the presence, for example, of triangles consisting of two parents and a child (with the two parent-child edges and one union edge). Because of the frequency of such cycles and the fact that they are the smallest possible cycles, we refer to them as \textit{trivial cycles}. The other typical \textit{familial cycle}, or cycle found within a family consisting of two parents and some number of children, is a cycle of length four consisting of two parents and two children.

Although familial cycles are ubiquitous in genealogical networks, they are not the only cycles that can form. Going far enough through an individual's ancestors, it is often possible to find a \textit{nearest common ancestor}, i.e., a common ancestor of one's father and mother. If such an ancestor exists (and it usually does exist), then the genealogical network has a nontrivial cycle. We refer to this as a \textit{common ancestor cycle}, which consists of only parent-child edges. Other nontrivial cycles are possible in genealogical networks via unions. For instance, a ``double cousins'' relationship occurs when two siblings from one family form unions with two siblings from another family. The result is a \textit{\horcycle}, or a cycle that contains only union edges and the parent-child edges connecting siblings. In genealogical networks, union and parent-child edges can combine in any number of ways to create complex non-tree structures (see Figure \ref{fig:1} left).

\begin{figure}[ht]
\begin{center}
    \begin{tabular}{c}
    \begin{overpic}[scale=0.55]{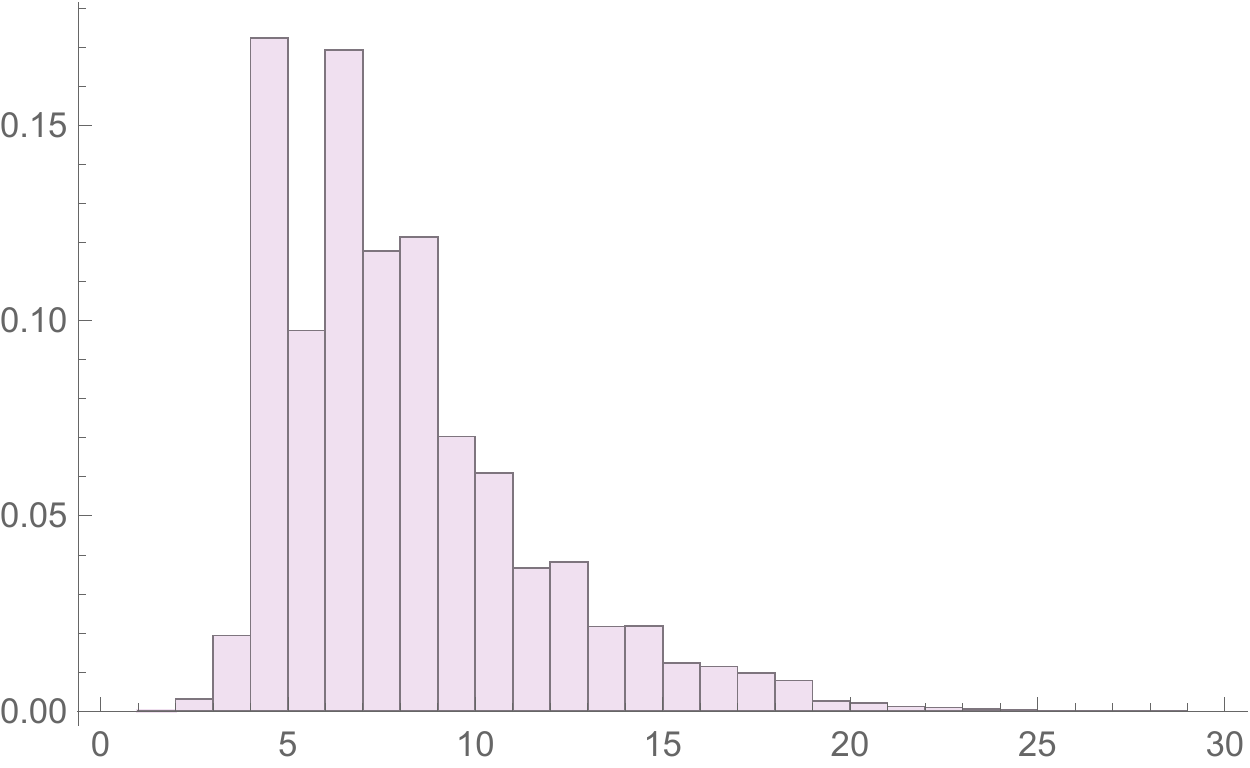}
    \put(28,-5){\small{Distance to Union}}
    \put(-10,4){\rotatebox{90}{\small{Fraction of Unions at Distance}}}
    \end{overpic}
\end{tabular}
\vspace{0.2cm}
\caption{The histogram representing the finite ``distance to union" distances is shown where data is collected from 104 genealogical networks from kinsources.net. The height of each bar represents the fraction of unions that form at a specific distance.}\label{fig:AggDistance}
\end{center}
\end{figure}

A feature that is particular to genealogical networks is that union edges typically form at specific distances within these networks. Here the distance $d(i,j)$ between $i$ and $j$ is the shortest path distance between these individuals if such a path exists. Otherwise, it is infinite. In a genealogical network we refer to the distance between two individuals before they form a union as the couple's \textit{distance to union}. For cultural, genetic, and other reasons these distance are typically not small, i.e. usually larger than four. Consequently, genealogical networks do not typically have small nonfamilial cycles and often have large extended cycles. This is illustrated in Figure \ref{fig:AggDistance} where distance to union data is collected from 104 publicly available genealogical networks given in Table \ref{Tab:2} in the Appendix. Here familial cycles are omitted and the height of each bar represents the fraction of unions that form at a specific distance. Noticeably, few unions form at distances less than five with the large majority of distance falling between 5 and 10.

\begin{figure}[ht]
\begin{center}
    \begin{tabular}{c}
    \begin{overpic}[scale=0.375]{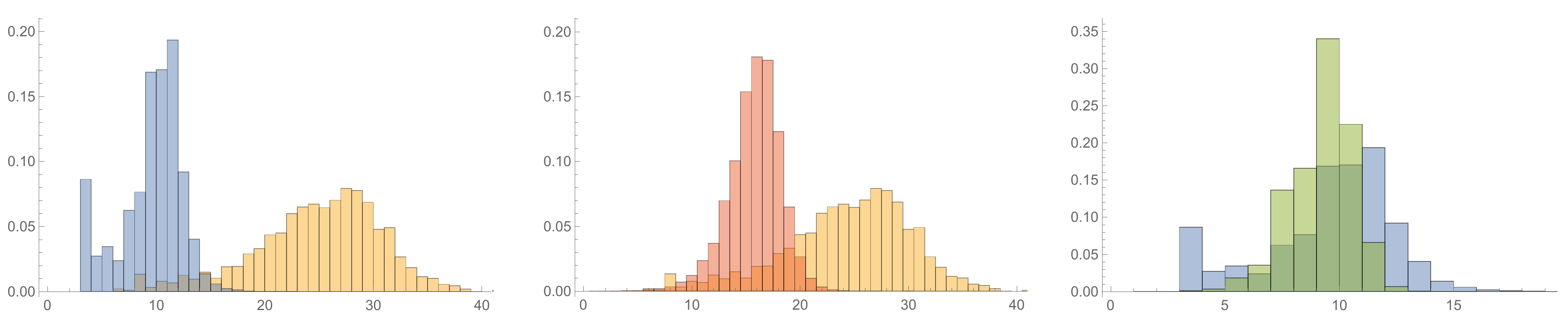}
    \put(4,-2.5){SM and DE Cycle Lengths}
    \put(39,-2.5){SM Configuration Model}
    \put(73,-2.5){DE Configuration Model}
    \end{overpic}
\end{tabular}
\vspace{0.25cm}
\caption{Left: Shown in orange is the distribution of the lengths of the cycles forming a basis of the nonfamilial cycle lengths in the San Marino (SM) genealogical network. The analogous distribution of cycle lengths is shown in blue for all cycles in the Deezer Europe (DE) social network. Center: Shown in orange is again the basis cycle length distribution of the San Marino genealogical network. In red is the distribution of the basis cycle lengths averaged over ten realizations of the (loopy, multi-edged) configuration model on the San Marino network. Since the configuration model generates graphs with the same degree distribution as the SM network, this panel indicates that SM's longer cycles do not arise simply from the degree distribution. Right: Shown in blue is again the basis cycle length distribution of the Deezer social network. In green is the distribution of the basis cycle lengths averaged over ten realizations of the configuration model on the Deezer social network. For this social network, the cycle length distribution can be mostly explained by the degree distribution alone.}\label{fig:famsocial}
\end{center}
\end{figure}

The observation that genealogical networks have large extended cycles is illustrated in Figure \ref{fig:famsocial}. Shown left in orange is the distribution of cycle lengths of the San Marino genealogical network, a network of the population of the Republic of San Marino from the 15th to the end of the 19th century~\cite{kin21}. In this network, which consists of 28,586 individuals, there are 7,146 familial cycles of length three and 8,636 familial cycles of length four. These are omitted in the figure so we can observe the lengths of the cycles forming a basis of nonfamilial cycles in the network. For the sake of contrast, in blue is the distribution of cycle lengths in a basis of the cycles found in the Deezer Europe social network, consisting of 28,281 individuals. Here, similar to genealogical networks, a social network is represented by a graph $G=(V,E)$ where the vertices $V$ also represent individuals. The difference is that in a social network the edges represent some type of social interaction(s). The Deezer network is an online music streaming platform whose social network represents individuals in Europe who use the platform where edges represent mutual user-follower relationships.

Noticeably, the San Marino network has relatively few nonfamilial basis cycles under length ten but quite a few cycles with lengths greater than thirty. In contrast, the Deezer social network has a much tighter distribution of basis cycles ranging from roughly five to fifteen in length.

To understand the extent to which these cycle distributions are related to the local structure of the associated networks we compare these to the cycle distribution of the associated configuration models of these two networks, respectively. The \textit{configuration model} is a model for generating random networks with a given degree sequence \cite{Newman2006}. Taking the degree sequences from both the San Marino genealogical and Deezer social network, we create ten versions of these networks each with the same degree sequences. The result of averaging the basis cycle length distributions of these versions of the San Marino and Deezer networks is shown in Figure \ref{fig:famsocial} (center and right in red and green, respectively). While the cycle distribution for the San Marino network is quite different from what the configuration model produces, the Deezer social network is quite similar to the distribution predicted by its configuration model. This suggests that much of the cycle structure in the Deezer social network is dominated by local interactions, whereas the cycles in the San Marino genealogical network are affected by nonlocal mechanisms that form the network. This includes, presumably, the nonlocal distance to union phenomena described above.

The relations we see in Figure \ref{fig:famsocial} between the cycle length distribution for the San Marino genealogical network and the Deezer social network are typical of the genealogical and social networks we consider in Section \ref{sec:5}. This suggests that cycle length distribution is a feature that can be used to distinguish genealogical from social networks. Specifically, when we consider two networks with a similar number of cycles, genealogical networks have a much wider distribution of cycle lengths than social networks. However, the method used to calculate the cycle length distribution in Figure \ref{fig:famsocial} does not provide any further insight into this phenomenon. This limitation motivates us to apply tools from persistent homology which provides ways to describe and measure the relation between \textit{any} two network cycles. The additional structure that can be obtained by these methods allow us to further distinguish the structure of genealogical and social networks (see Section \ref{subsec:apply_bottleneck_distance}) and to relate the structural differences demonstrated in Figure \ref{fig:famsocial} to mechanisms that produce genealogical and social networks, respectively (see Section \ref{subsec:connections}).

\section{Persistent Homology of Networks}\label{sec:3}
Persistent homology provides a method for studying cycles in a network. For the purposes of this paper, a brief explanation of persistent homology will be given from the context of simplicial homology. For a more in-depth treatment of simplicial homology, see Chapter 2.1 of \cite{hatcher}. For those readers who are either familiar with the basics of persistent homology or who wish to skip the following technical discussion it is possible to proceed to Section 5 where we discuss the social and genealogical networks we analyze.

For a network given by a graph $G=(V,E)$ we define the distance matrix $D(G)=[d_{ij}]$ to have entries $d_{ij}=d(i,j)$, which is the length of the shortest path between individual $i$ and $j$. For each value $\delta$ that appears in the distance matrix $D(G)$, we form a simplicial complex $G_{\delta}$ as follows. The set of $0$-simplices is equivalent to the set of vertices of $G$, where each $0$-simplex is identified with a single vertex. Since the distinction between $0$-simplices and vertices is purely formal, we will use the terms $0$-simplex and vertex interchangeably, and the $0$-simplices will be indexed the same way as the vertices. The set of $1$-simplices $E_{\delta}$ corresponds to the set of edges $\edge{i}{j}$ such that $d(i,j) \leq \delta$, where the edge $\edge{i}{j}$ is identified with the $1$-simplex formed by $\vertex{i}$ and $\vertex{j}$. Again the distinction here is unnecessary for our present discussion, so we will use the same notation for $1$-simplices and edges. However, the simplicial complex $G_{\delta}$ may also contain objects that do not have equivalent representatives in the graph $G$, namely the $n$-simplices for $n\geq 2$. For each integer $n \geq 2$, the set of $n$-simplices in $G_{\delta}$ consists of all $n$-simplices $\bigsimplex{a_0&a_1&\ldots&a_n}$ such that $d(a_i,a_j) \leq \delta$ for $0\leq i < j \leq n$. That is, $G_{\delta}$ includes an $n$-simplex $\sigma$ if each vertex listed in $\sigma$ is within $\delta$ of every vertex listed in $\sigma$.

In order to simplify our remaining definitions, we extend our definition of $G_{\delta}$ to include all non-negative integers. For $i\geq 0$, let $\delta_i$ be the greatest entry of $D(G)$ such that $\delta_i \leq i$. Let $G_i = G_{\delta_i}$. This definition together with our construction of $G_\delta$ ensures the following three important properties are true for all $G_i$.
\begin{enumerate}
    \item For $i < j$, $G_{i}$ is a subcomplex of $G_{j}$, i.e. every simplex of $G_{i}$ is a simplex of $G_{j}$.
    \item For $i \geq 1$, there exists a subcomplex of $G_{i}$ that can be identified with the original graph $G$.
    \item Since $G$ is finite, let $M=\max_{ij}d(i,j)$, then, for all $i \geq M$, $G_{i} = G_M$.
\end{enumerate}

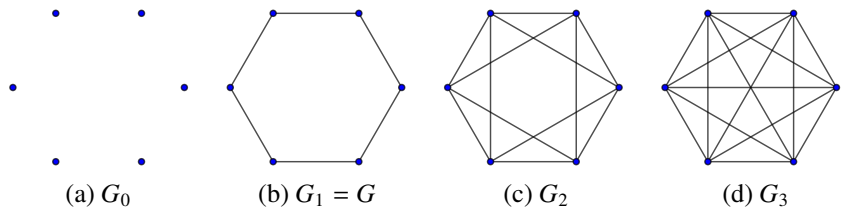
\begin{figure}[ht]
\begin{center}
\begin{tabular}{c}
    \begin{overpic}[scale=0.55]{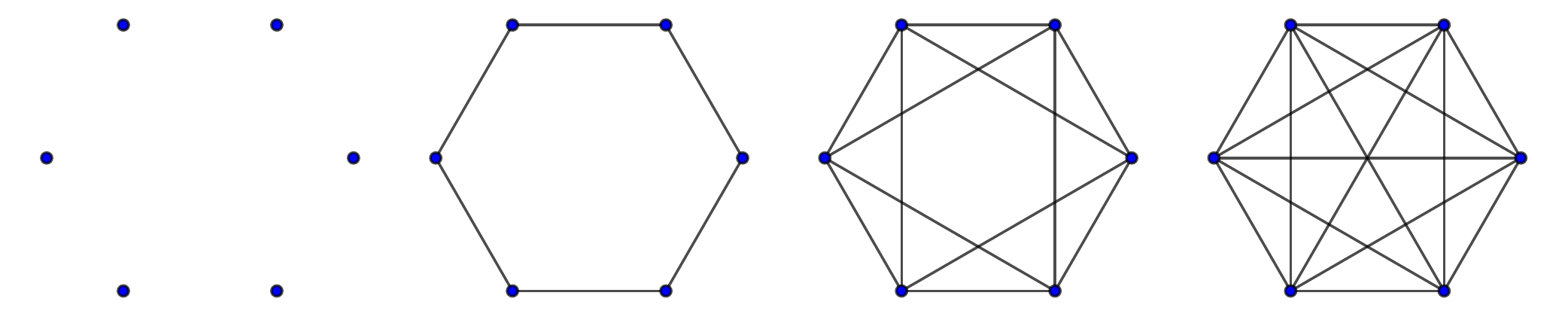}
    \put(9,-3){(a) $G_0$}
    \put(31,-3){(b) $G_1=G$}
    \put(59,-3){(c) $G_2$}
    \put(84,-3){(d) $G_3$}
    \end{overpic}
\end{tabular}
\vspace{0.25cm}
\caption{The hexagonal network $G=G_1$ in Example \ref{ex:six_cycle_G_i} is filled in as $i$ increases from $0$ to $3$. This produces the simplicial complexes $G_0, G_1, G_2, G_3$ shown left to right.}\label{fig:hex1}
\end{center}
\end{figure}

\begin{example} \textbf{(Hexagonal Network)}
\label{ex:six_cycle_G_i}
Consider the \textit{hexagonal network} $G=(V,E)$ with six vertices, forming a single cycle, shown in Figure \ref{fig:hex1}(b). This network has the distance matrix
\[D(G) = \begin{bmatrix}
0 & 1 & 2 & 3 & 2 & 1\\
1 & 0 & 1 & 2 & 3 & 2\\
2 & 1 & 0 & 1 & 2 & 3\\
3 & 2 & 1 & 0 & 1 & 2\\
2 & 3 & 2 & 1 & 0 & 1\\
1 & 2 & 3 & 2 & 1 & 0\end{bmatrix}.\]
For the values $i=0$, $1$, $2$, $3$, we form four simplicial complexes, $G_0$, $G_1$, $G_2$, and $G_3$ where we let $G_i=(V_i,E_i)$. For $i=0$, $E_0$ is empty. Thus, $G_0$ consists of six vertices. For $i=1$ the set $E_1$ contains the six edges that form the network's single cycle, so $G_1=G$. This graph has no \textit{trivial cycles} (i.e., triangles), so $G_1$ contains no simplices of \textit{dimension} greater than 1 (i.e., no $n$-simplices for $n>1$). For $i=2$ the set $E_2$ gains six additional edges. We also now have eight trivial cycles. Each of these cycles is the boundary of a 2-simplex, so $G_2$ contains these eight 2-simplices as well. However, no subset of these 2-simplices forms the boundary of a 3-simplex, so $G_2$ has no simplices of dimension greater than 2. For $i=3$ the set $E_3$ contains all possible edges between the vertices of $G$, so all possible trivial cycles are present. Additionally, all possible 2-simplices, and hence all possible $n$-simplices, are also present in $G_3$. In particular, $G_3$ is a 6-simplex with its boundary. Since $M=3$ is the largest value we see in the distance matrix, then $G_i = G_3$ for $i \in \ZZ$, $i > 3$.
\end{example}



The persistent homology of the network $G$ measures how the homology of $G_{i}$ changes as $i$ increases. If certain features can be identified across multiple values of $i$, we say they \textit{persist}. Intuitively, features that arise from the actual network structure should persist for many values of $i$, while features that arise because of measurement error, `noise', should only appear sporadically. The Stability Theorem (the Main Theorem of \cite{stability}) states that if the error in measuring a network is bounded by some constant $C$, then the persistent homology of the true network and the persistent homology of the noisy network will differ by \textit{at most} $C$. We will make this statement more precise in Section \ref{subsec:bottleneck}.

Here we give a formal definition of persistent homology in terms of simplicial homology, which we will immediately follow this with equivalent definitions in the context of networks. We use $H_p(G_i)$ to denote the dimension-$p$ simplicial homology of the simplicial complex $G_i$ with coefficients in $\ZZ_2$, as $H_p(X)$ is a vector space of $\ZZ_2$.

\begin{definition}\textbf{(pth Persistent Homology)}
For a graph $G$, and integers $i,j$ with $0\leq i \leq j$, let the function $\phi_{i,j} : H_p(G_i) \to H_p(G_j)$ be the linear map induced by the inclusion $G_i\to G_j$. The $p$th persistent homology of $G$, $PH_p(G)$ is the pair $(\{H_p(G_i)\}_{ i \ge 0},\{\phi_{i,j}\}_{0 \leq i < j})$.
\end{definition}

Our analysis in Sections \ref{sec:4}-\ref{sec:results} only requires the first few dimensions of persistent homology to distinguish the genealogical and social networks we consider. In order to better understand what persistent homology calculates, in what follows we will provide equivalent definitions for $PH_0$, $PH_1$, and $PH_2$ using network concepts. We also illustrate how these definitions apply to the hexagonal network in Figure \ref{fig:hex1}(b). (See Examples \ref{ex:six_cycle_PH0}, \ref{ex:six_cycle_PH1}, and \ref{ex:six_cycle_PH2} for $PH_0$, $PH_1$, and $PH_2$; respectively.)

\begin{definition}\label{def:BD}\textbf{(Births and Deaths)}
Let $G=(V,E)$ be a network with simplicial complexes $G_0, G_1, G_2, \cdots$. The $p$th persistent homology of $G$ provides maps $\phi_{i,j}$ between the $p$th homology of $G_i$ and the $p$th homology of $G_j$. Suppose that basis elements have been chosen for each $H_p(G_i)$ so that if $\alpha$ is a basis element of $H_p(G_i)$, then $\phi_{i,j}(\alpha)$ is either trivial in $H_p(G_j)$ or a basis element of $H_p(G_j)$. The birth of a basis element $\alpha \in H_p(G_j)$ is the minimum index $i$ such that $\alpha = \phi_{i,j}(\hat{\alpha})$ for some basis element $\hat{\alpha} \in G_i$. The death of $\alpha$ is the minimum index $k$ such that $\phi_{j,k}(\alpha)$ is trivial.
\end{definition}

\begin{remark} Those already familiar with persistent homology will find that the preceding definition is somewhat nonstandard, although it is equivalent to the standard definition. We have taken this approach to reduce the notation burden on non-specialist readers. We have done similarly with some of the other persistent homology definitions. 
\end{remark}

We will demonstrate how to choose such representatives for $H_0$, $H_1$, and $H_2$ in the following definitions. Given such representatives, though, the maps $\phi_{i,j}$ and $\phi_{j,k}$ are simply the maps on homology induced by the inclusion maps $G_i \subset G_j \subset G_k$. That is, if $a$ represents $\alpha \in H_p(G_i)$, then $a$ also represents $\phi_{i,j}(\alpha)$. The Fundamental Theorem of Persistent Homology ensures that we can choose a single representative that corresponds to $\alpha \in H_p(G_j)$, $\hat{\alpha} \in H_p(G_i)$, and $\phi_{j,k}(\alpha) \in H_p(G_k)$. The birth of $\alpha$ is then just the first $G_i$ in which the representative exists, and the death of $\alpha$ is the first $G_k$ in which the representative is \textit{null-homotopic} i.e., homotopic to a trivial cycle.

\begin{definition}\textbf{(Representing Persistent Homology: Dimension 0)}
\label{def:Rep0}
Let $G = (V,E)$ be a network with vertices $V = \{1,2,\ldots,n\}$ which form $k$ connected components. Then $H_0(G_0) \cong \ZZ_2^n$, so we can identify the basis for $H_0(G_0)$ with the set of all $n$ vertices. Likewise, we may choose $k$ vertices, one from each connected component, to represent the basis for $H_0(G_i) \cong \ZZ_2^k$ for $i \geq 1$. Thus, we will refer to the vertices of $G$ as \textit{\PHreps} of $PH_0(G)$. (In fact, $PH_0(G)$ is a vector space whose basis elements are equivalence classes of formal sums of $0$-simplices.)
\end{definition}

\begin{example}
\label{ex:six_cycle_PH0}
We now consider $PH_0(G)$ for the hexagonal network $G$ in Figure \ref{fig:hex1}, with $G_0$, $G_1$, $G_2$, and $G_3$ in the same figure. Recall that $G$ has six distinct vertices forming one connected component. If we take any numbering of the vertices, $V=\{1,2,3,4,5,6\}$, then $H_0(G_0) \cong \ZZ_2^6$, which is equivalent to the vector space over $\ZZ_2$ with basis $V$. For $i>0$, $H_0(G_i) \cong \ZZ_2$, which is equivalent to the vector space over $\ZZ_2$ with basis $\{1\}$. For any $v \in V$, since $i=0$ is the first time we see $v$, we call this the \emph{birth} of $v$. At $i=1$, since we have removed all vertices except $1$ from the basis, we say this is the \emph{death} of those five 0-simplices. Since $1$ will always be in the basis for $G_i$, the \emph{death} of $1$ is said to be $\infty$.
\end{example}

\begin{definition}\textbf{(Representing Persistent Homology: Dimension 1)}
\label{def:Rep1}
Let $G = (V,E)$ be a network with one connected component. For each $i \geq 0$, we can identify the basis of $H_1(G_i)$ with a set $C_i$ of cycles in $G_i$. The Fundamental Theorem of Persistent Homology allows us to choose these cycles so that if $\sigma$ is a cycle in $C_i$, then exactly one of the following is true for any integer $j\geq 0$:
\begin{enumerate}
    \item $\sigma$ does not exist in $G_j$, in which case $j<i$,
    \item $\sigma$ is trivial or null-homotopic in $G_j$, in which case $i<j$,
    \item $\sigma$ is a cycle in $C_j$.
\end{enumerate}
Thus, we will refer to the cycles in $\bigcup_{i\geq 0} C_i$ as the \textit{\PHreps} of $PH_1(G)$. (Again, $PH_1(G)$ is actually much larger than this. These are actually representatives of equivalence classes that form a basis for $PH_1(G)$ as a vector space.)
\end{definition}

We note that $C_0$ is always empty, since there are no edges in $G_0$. Furthermore, $\operatorname{rank}(H_1(G_i)) = |C_i|$ for all $i \geq 0$. Because of the construction of the $G_i$ all \PHreps\ of $PH_1(G)$ will be present in $G_1$. One can think of the \PHreps\ of $PH_1(G)$ as representing ``large" cycles. More specifically, if a cycle $\sigma$ is contained in $\bigcap_{s \leq i \leq t} C_i$, then it must have a diameter of at least $t$ and at least one pair of consecutive vertices distance $s$ apart.

\begin{figure}[ht]
    \begin{tabular}{cccc}
    \includegraphics[page=1,scale=1]{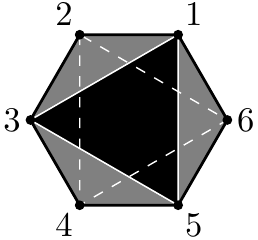}
    &
    \includegraphics[page=2,scale=1]{Fig5Remake.pdf}
    &
    \includegraphics[page=3,scale=1]{Fig5Remake.pdf}
    &
    \includegraphics[page=4,scale=1]{Fig5Remake.pdf}\\
    (a) $G_2$ trivial cycles &
    (b) $G_2$ null-homotopic cycle &
    (c) $G_2$ sphere &
    (d) $G_3$ select 3-simplices
    \end{tabular}
 \caption{ A visual depiction of simplices and cycles present in $G_2$. Left: Four trivial cycles filled by individual $2$-simplices: $\triad{1}{2}{3}$, $\triad{3}{4}{5}$, $\triad{1}{5}{6}$, and $\triad{1}{3}{5}$. Center Left: A non-trivial, but null-homotopic cycle, {$1,2,3,5,1$} filled in by two $2$-simplices $\triad{1}{2}{3}$ and $\triad{1}{3}{5}$. Center Right: All eight $2$-simplices  represented as the faces of a regular octahedron. Right: The closed surface of $G_2$ is filled in by four $3$-simplices $\tetrahedron{1}{2}{3}{6}, \tetrahedron{1}{3}{5}{6} (not shown), \tetrahedron{3}{4}{5}{6}, \tetrahedron{2}{3}{4}{6}$.}
\label{fig:filled_in_cycle} \label{fig:t2flat} \label{fig:t2nullhomotopy} \label{fig:t2filled3d}
\end{figure}

\begin{example}
\label{ex:six_cycle_PH1}
We now consider $PH_1(G)$ for the hexagonal network $G$ in Figure \ref{fig:hex1}(b). In both Figure \ref{fig:hex1}(a) and \ref{fig:hex1}(b) we see that $G_0$ has no cycles, $G_1$ has exactly one cycle, and that the cycle in $G_1$ is non-trivial. In Figures \ref{fig:t2flat}(a) and \ref{fig:t2nullhomotopy}(b), we have indicated some of the cycles in $G_2$, namely the cycles 1,2,3,1; 3,4,5,3; 1,5,6,1; and 1,3,5,1 in Figure \ref{fig:t2flat}(a) and the cycle 1,2,3,5,1 in Figure \ref{fig:t2nullhomotopy}(b). In fact, Figure \ref{fig:t2filled3d}(c) shows us that $G_2$ is an octahedron and therefore every cycle in $G_2$ is either trivial or null-homotopic. Finally, $G_3$ contains even more cycles than $G_2$, such as 1,3,6,1; but these are all null-homotopic since $G_3$ also contains every possible $2$-simplex for six vertices. Therefore, $PH_1(G)$ has only one \PHrep, the cycle 1,2,3,4,5,6,1; which appears in $G_1$, so we say that $t=1$ is the \emph{birth} of the cycle. The cycle is null-homotopic in $G_2$, so $t=2$ is the \emph{death} of the cycle.
\end{example}

We now turn our attention to $PH_2(G)$, but in order to represent $PH_2(G)$ we need to introduce some new structure for the induced graphs. A \textit{triangle} $\triad{a}{b}{c}$ in $G_i$ is a set of three vertices, $a$, $b$, and $c$, that form a trivial cycle in $G_i$. That is, the edges $\edge{a}{b}$, $\edge{b}{c}$, and $\edge{a}{c}$ are all present in $G_i$. A \textit{closed surface} in $G_i$ is a set of distinct triangles so that for each $\triad{a}{b}{c}$ in the set there is exactly one other triangle $\triad{a}{b}{d}$ also in the set.  A closed surface in $G_i$ is \textit{trivial} if the corresponding set of $2$-simplices is null-homotopic in $G_i$. That is, the closed surface is ``filled in'' by some collection of $3$-simplices in $G_i$. For example, the octahedron in Figure \ref{fig:filled_in_cycle}(c) is a non-trivial closed surface in $G_2$ because there are no 3-simplices in $G_2$. In $G_3$, however, we add edges between vertices at distance 3. In turn, we gain several 3-simplices, including $\tetrahedron{1}{2}{3}{6}$, $\tetrahedron{1}{3}{5}{6}$, $\tetrahedron{3}{4}{5}{6}$, and $\tetrahedron{2}{3}{4}{6}$. Figure \ref{fig:filled_in_cycle}(d) shows three of these 3-simplices to demonstrate how the closed surface from $G_2$ is filled in by all four.\\

\begin{definition}\textbf{(Representing Persistent Homology: Dimension 2)}
\label{def:Rep2}
Let $G = (V,E)$ be a network with one connected component. For each $i \geq 0$, we can identify the basis for $H_2(G_i)$ with a set $S_i$ of non-trivial closed surfaces in $G_i$. The Fundamental Theorem of Persistent Homology allows us to choose these representatives so that if $\sigma$ is a closed surface in $S_i$, then exactly one of the following is true for any integer $j \geq 0$
\begin{enumerate}
    \item $\sigma$ does not exist in $G_j$, in which case $j<i$,
    \item $\sigma$ is trivial in $G_j$, in which case $i<j$,
    \item $\sigma$ is a cycle in $S_j$.
\end{enumerate}
Thus we will refer to the closed surfaces in $\bigcup_{i\geq 0} S_i$ as the \textit{\PHreps} of $PH_2(G)$.
\end{definition}

The geometric intuition for $PH_2(G)$ is similar to that of  $PH_1(G)$ in identifying large `voids' in $G$. If $\sigma \in \bigcap_{s \leq i \leq t} S_i$, then $\sigma$ is a closed surface with diameter at least $t$. The value of $s$ is harder to describe, but is related to the density of vertices.

\begin{example}
\label{ex:six_cycle_PH2}
We now consider $PH_2(G)$ for the hexagonal graph $G$ in Example \ref{ex:six_cycle_G_i}. Recall from Example \ref{ex:six_cycle_PH1} that $G_0$ and $G_1$ have no trivial cycles, and therefore contain no closed surfaces. We can see in Figure \ref{fig:t2filled3d} that $G_2$ has exactly one closed surface and it must be non-trivial, since there are no 3-simplices. Finally, $G_3$ has many closed surfaces, but because it contains every possible $3$-simplex on six vertices, these are all trivial. Therefore, $PH_2(G)$ has only one \PHrep, the octahedral closed surface in $G_2$. This surface first appears in $G_2$, so $t=2$ is its \emph{birth}, and the surface is filled by a solid in $G_3$, so $t=3$ is its \emph{death}.
\end{example}

\begin{definition}\textbf{(Persistence Intervals)}
\label{def:PHIntervals}
Recall that the birth of a \PHrep\ $\mathbf{\sigma}\in PH_p(G)$ (vertex, cycle, or closed surface) of the persistent homology of a network $G$ is the smallest integer $i$ so that $\sigma \in G_i$, and the \textit{death} of $\sigma$ is the largest integer $j$ so that $\sigma \in G_{j-1}$ and $\sigma$ is trivial in $G_k$ for $k \geq j$, if such an integer exists. The \emph{persistence interval} for $\sigma$ is $\Pinterval{a}{b}$, where $a$ and $b$ are the birth and death of $\sigma$, respectively. This represents the set of all parameter values $i$ for which the equivalence class corresponding to $\sigma$ is a non-trivial element of $H_p(G_i)$. The \textit{persistence} of $\sigma$ is $b-a$.
\end{definition}

\begin{example}
\label{ex:six_cycle_persistence}
We now finish our consideration of the persistent homology of $G$ from Figure \ref{fig:hex1}(b). Recall from Example \ref{ex:six_cycle_PH0} that $PH_0(G)$ has six \PHreps. These all have birth $t=0$. Five of these have a death of $t=1$, and one of these has a death of $\infty$. Therefore the persistence intervals for $PH_0(G)$ are $\Pinterval{0}{1} \times 5$ and $\Pinterval{0}{\infty} \times 1$.

From Example \ref{ex:six_cycle_PH1}, we know $PH_1(G)$ has one \PHrep, with birth $t=1$ and death $t=2$. Therefore the corresponding persistence interval is $\Pinterval{1}{2}$. Note that the diameter of the cycle is $3$ and every pair of consecutive vertices is distance $1$ apart. This follows the idea mentioned earlier that the \PHreps\ of $PH_1(G)$ indicate `large' cycles. Specifically, the diameter of $\sigma$ is \textit{at least} the death of $\sigma$, and the birth of $\sigma$ is the maximum distance between consecutive vertices.

From Example \ref{ex:six_cycle_PH2}, $PH_2(G)$ has one \PHrep, with birth $t=2$ and death $t=3$. Therefore, the persistence interval for that element is $\Pinterval{2}{3}$. Note that the diameter of the corresponding set of vertices is $3$ in $G$. This also follows the idea mentioned earlier that $PH_2(G)$ identifies large `voids' in $G$. Specifically, the death of $\sigma$ is a lower bound on the diameter of $\sigma$.
\end{example}

Given the representatives chosen in Definitions \ref{def:Rep0}, \ref{def:Rep1}, \ref{def:Rep2}, and \ref{def:PHIntervals}, we have the following three observations regarding the persistent homology of a finite, undirected, unweighted graph $G$:\\
\textbf{(i)} If $G$ has $n$ vertices, then $PH_0(G)$ will have exactly $n$ persistence intervals, with exactly one $\Pinterval{0}{\infty}$ interval for each connected component and the rest will be $\Pinterval{0}{1}$ intervals.\\
\textbf{(ii)} In dimension 1, $PH_1(G)$ describes the number and sizes of the non-trivial cycles in the original network. The persistence intervals will all be of the form $\Pinterval{1}{b}$ for some integer $b>1$. The value of $b$ is related to the diameter of the corresponding cycle. In the networks we have studied, we note that a persistence interval $\Pinterval{1}{b}$ in $PH_1(G)$ corresponds to a simple cycle with between $3b-2$ and $3b$ vertices, inclusive.\\
\textbf{(iii)} In dimension 2, the voids we detect in $PH_2(G)$ tell us about the nontrivial intersections of cycles. Such intersections are hard to visualize but, roughly speaking, a \PHrep\ in $PH_2(G)$ can only form if several large cycles intersect each other pairwise.

\section{Comparing Networks using Persistent Homology}\label{sec:4}
In this section we demonstrate how methods based on persistent homology can be used to compare different networks. The two methods we introduce in this paper are based on using (a) the bottleneck distance and (b) the persistence curves of a given set of networks. Both (a) and (b) rely on first computing persistence intervals then analyzing the differences in these intervals.

The two networks we consider throughout this section to demonstrate these methods are the Tikopia genealogical network from Figure \ref{fig:1} (left) and the hexagonal network from Figure \ref{fig:hex1}. The persistence intervals for these networks are given in Table \ref{table:Pintervals}, respectively.


\begin{table*}[ht]
\centering
\begin{tabular}{l | l | l}
\hline
\hline
\multirow{2}{*}{Dimension} & \multicolumn{2}{c}{Interval Type and Persistence}\\
\cline{2-3}
& Tikopia & Hexagon\\
\hline
Dimension 0  & $[0,\infty)\times8, \ [0,1)\times286$ & $[0,\infty)\times1, \ [0,1)\times1$\\
\hline
\multirow{2}{*}{Dimension 1}  & \multirow{2}{18em}{$[1,2)\times16$,  $[1,3)\times19$, $[1,4)\times5$, $[1,5)\times3$, $[1,6)\times2$, $[1,7)\times1$} & \multirow{2}{*}{$[1,2)\times1$}\\
&&\\
\hline
\multirow{2}{*}{Dimension 2}  & \multirow{2}{18em}{$[2,3)\times4$, $ [3,4)\times11$, $[4,5)\times12$, $ [5,6)\times4$, $[6,7)\times5$, $[7,8)\times1$, $[8,9)\times1$} & \multirow{2}{*}{$[2,3)\times1$}\\ &&\\
\hline
\hline
\end{tabular}
\vspace{.1cm}
\caption{The persistence intervals of the Tikopia genealogical network and the hexagon network are shown. Here the notation $[a,b)\times k$ indicates that the network has $k$ persistence intervals $[a,b)$. The corresponding persistence diagrams are shown in Figure \ref{fig:PD} and the corresponding persistence curve for the Tikopia network is shown in Figure \ref{fig:persistence_curve}.}\label{table:Pintervals}
\end{table*}


\subsection{Persistence Diagrams and Bottleneck Distance}
\label{subsec:bottleneck}
One common way to represent persistence intervals is to plot them as points in $\R \times (\R\cup \{\infty\})$, which is typically referred to as a persistence diagram. While this method of visualizing a network's persistent homology does not indicate how often a given persistence interval occurs, it does provide information on what kind of persistence intervals occur for a given network.

\begin{definition}\textbf{(Persistence Diagrams)}
Let $PH_p(G)$ be the $p$th persistent homology of a network $G$. The persistence diagram for $PH_p(G)$ is a multiset of points in $\R \times (\R\cup \{\infty\})$ defined as follows.
\begin{itemize}
    \item For each $\sigma \in PH_p(G)$ with persistence interval $\Pinterval{a}{b}$, we include one copy of the point $(a,b)$.
    \item For each $c \in \R$, we include infinitely many copies of the point $(c,c)$.
\end{itemize}
\end{definition}

Note that we include the points $(a,a)$ to represent features in $G$ that are considered trivial in $PH_p(G)$, such as cycles consisting of exactly three vertices. This inclusion is necessary for us to define a meaningful metric on the space of persistence diagrams. The metric we use here is called the bottleneck distance.

\begin{definition}\textbf{(Bottleneck Distance)}\label{def:bn}
 Let $S_1$ and $S_2$ be persistence diagrams for two graphs $G$ and $H$, respectively. Let $\eta$ range over the set of bijections from $S_1$ to $S_2$. Then the \emph{bottleneck distance} between $S_1$ and $S_2$ is
 \[d_B(S_1, S_2) = \inf_{\eta} \sup_{x \in S_1} \| x - \eta(x)\|_{\infty}.\]
\end{definition}



The Fundamental Theorem of Persistent Homology (introduced in \cite{Zomorodian2005}, explained well in \cite{Otter17} and \cite{AAF19}) ensures that if two graphs are isomorphic, the corresponding persistence diagrams will be equal, and thus the bottleneck distance will be 0. However, it is possible for non-isomorphic graphs to have identical persistence diagrams.

\begin{figure}[ht]
\begin{center}
    \begin{tabular}{c}
    \begin{overpic}[scale=0.375]{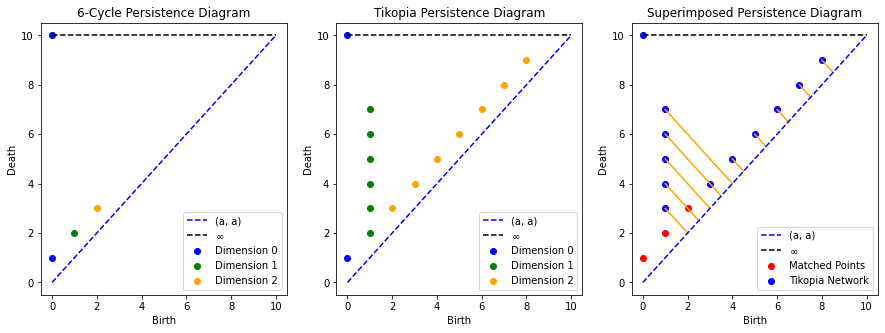}
    \put(4,-2.5){Hexagonal Network PD}
    \put(39,-2.5){Tikopia Network PD}
    \put(73,-2.5){Bottleneck Bijection}
    \end{overpic}
\end{tabular}
\vspace{0.25cm}
\caption{Left: The persistence diagram of the hexagonal network in Figure \ref{fig:hex1}(b) is shown. Center: The persistence diagram of the Tikopia genealogical network in Figure \ref{fig:1} (left) is shown. Right: A bottleneck bijection between the persistence intervals of the hexagonal and Tikopia family network is shown. Orange lines show which points are matched to points of the form $(a,a)$ where $a \in \mathbb{R}$.}\label{fig:PD}
\end{center}
\end{figure}

\begin{example} \textbf{(Bottleneck Distance Between the Tikopia and Hexagonal Networks)}
\label{ex:bottleneck_example}
Notice that the persistence intervals for the Tikopia genealogical network (see Table \ref{table:Pintervals}) include, as a subset, the persistence intervals from the hexagonal network we considered in Example \ref{ex:six_cycle_persistence}. We can form a bijection between the persistence diagrams of the Tikopia and hexagonal network by identifying the non-trivial intervals from the hexagonal network with those of the Tikopia network.
We then map any additional intervals from the Tikopia network of the form $\Pinterval{a}{b}$ to the trivial interval $\Pinterval{\frac{a+b}{2}}{\frac{a+b}{2}}$. (The perceptive reader may notice that this is not clearly a bijection, but there is a standard technique from set theory for modifying it to be bijective.)

This mapping is shown in Figure \ref{fig:PD} (right). Here, $\Pinterval{1}{7}$ is mapped to $\Pinterval{4}{4}$. As this pair of points is further apart than any other pair in this bijection, the bottleneck distance for the two networks is \emph{at most three}, since we take an infimum over all possible bijections. Conversely, there is no interval in the hexagonal persistence diagram that is closer to $\Pinterval{1}{7}$ than 3, so the bottleneck distance is \emph{at least three}. Thus, the bottleneck distance for these two persistence diagrams is exactly 3.
\end{example}

Suppose that two networks, each of which is connected, admit isometric embeddings in $\R^n$. The Stability Theorem \cite{stability} guarantees that if the Hausdorff distance between the embeddings is $\delta$, then the bottleneck distance for the corresponding persistence diagrams is at most $\delta$. For example, if the $PH_1$ persistence diagrams differ by $\delta$, then any attempt to pair up cycles in the networks must include at least one pair of cycles for any isometric embedding that are $\delta$ apart in that embedding. In Section \ref{subsec:apply_bottleneck_distance} we apply this idea to a large collection of genealogical and social networks.


\subsection{Persistence Curves}
For the network data we consider, persistence diagrams obfuscate a key difference that we consider important: the number of persistence intervals. For a simple example of this, consider networks of the form $V = \{1,2,\ldots,n\}$ with edges of the form $\edge{i}{i+1}$ for $1 \leq i < n$. For $n \geq 2$, any network of this type will have persistence intervals $\Pinterval{0}{1}\times (n-1)$ and $\Pinterval{0}{\infty}\times 1$. However, when plotting the persistence diagram we will only `see' two points: $(0,1)$ and $(0,\infty)$.

\begin{figure}[ht]
\begin{center}
    \begin{tabular}{cc}
    \begin{overpic}[scale=0.44]{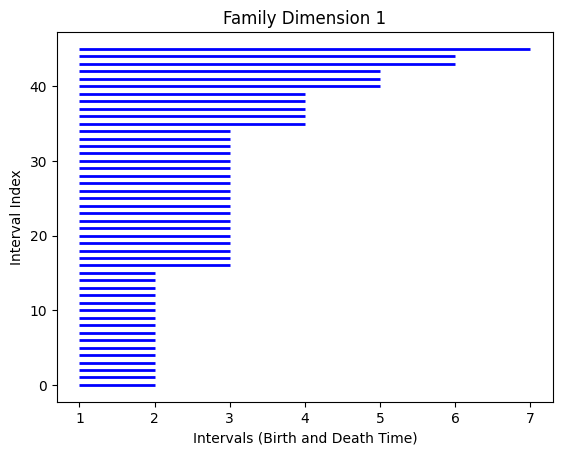}
    \put(24,-5){Tikopia Network Barcode}
    \end{overpic}
    \hspace{1cm}
    \begin{overpic}[scale=0.44]{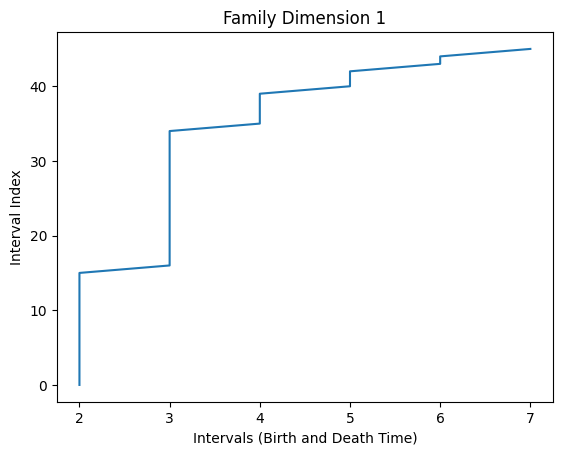}
    \put(14,-5){Tikopia Network Persistence Curve}
    \end{overpic}
\end{tabular}
\vspace{0.25cm}
\caption{Left: The barcode of the Tikopia genealogical network in dimension 1 is shown. The individual bars are formed from the persistence intervals given in Table \ref{table:Pintervals}. Right: The associated persistence curve for the Tikopia network in Figure \ref{fig:1} is shown.}\label{fig:persistence_curve}
\end{center}
\end{figure}

To address this limitation, we introduce the notion of a \textit{persistence curve} as a new way to visualize the persistent homology of a network (see Definition \ref{def:percurv}). The difference between the persistence curve and the persistence diagram of a network is that the persistence curve also includes the number of intervals of a particular type. To create a persistence curve we first compute a network's persistence intervals, then sort the intervals of a given dimension by their persistence into a bar graph. For instance, in dimension 1 the Tikopia genealogical network has thirteen $[1,2)$ intervals, nineteen $[1,3)$ intervals, etc. which are sequentially stacked as shown in Figure \ref{fig:persistence_curve} (left) to create what we will call a \textit{barcode}. To create the associated persistence curve we connect the endpoints of each subsequent bar as shown in Figure \ref{fig:persistence_curve} (right).

In dimension-one, the birth times of our intervals will all start at 1, as the networks we consider are unweighted, undirected, and connected. This means that in this dimension the resulting bar graph is also a plot of the death times for each interval. For higher-dimensions, which have varied birth times, we also plot the lengths of the intervals but for simplicity we start at 1 as in dimension-one.

A formal definition of a network's persistence curves is the following.

\begin{definition}\textbf{(Persistence Curves)}\label{def:percurv}
Let $G=(V,E)$ be a network with nonempty vertex and edge sets. Let $\{[a_j,b_j)\}$ be the set of all persistence intervals for each $\sigma_{j} \in PH_n(G)$ where $j \in \NN$. For all $n \in \NN$ the persistence curve  $PH_n(G)$ is the linear interpolation of the set of points $\{(b_j - (a_j - 1), j)\}$ where $b_{j-1} - (a_{j-1} - 1) \leq b_j - (a_j - 1)$.
\end{definition}

Visualizing persistence intervals as a curve allows us to compare the persistent homology of different networks in a similar fashion to persistence diagrams while retaining different information.  In particular, we can see how many intervals there are of a given persistence, whereas the persistence diagram only indicates the presence of such an interval. In what follows we will typically plot the persistence curves of multiple networks on the same axes to indicate what differences exist in the persistent homology of different networks (cf. Section \ref{sec:results}).

\section{Data}\label{sec:5}
The data we consider in this paper is of two types; genealogical network data and other social network data. The genealogical networks we consider are drawn from ninety-seven genealogical networks found in\cite{kin21}, which range in size from $n=17$ to $5,016$ individuals. The social network data we use is taken from twenty-seven different social networks obtained from \cite{kon21,snap21,rep21,vlad21}. These range in size from $n=16$ to $2,539$ individuals. (See Table \ref{Tab:2} in the Appendix for a full description of this data set.)

Although many larger genealogical and social network data sets are available we are limited by both the temporal and spacial complexity of the algorithm used to compute persistence intervals. The program we used, called \textit{Ripser} (from the python package Ripser) \cite{ripser}, has a computational and spacial complexity of $O((n+m)^3)$ where $n$ is the number of individuals and $m$ is the number of edges in a network. The number $n+m$ is the number of simplicies in the network. In the genealogical networks we consider there are between $n+m=41$ to $15,735$ simplicies and in the social networks we consider between $n+m=41$ to $19,056$ simplices.

To understand how a network's persistence intervals are effected by the completeness or incompleteness of data we also consider subnetworks sampled from a few, much larger, genealogical and social networks. These \textit{sampled networks} are created by randomly selecting an individual with a single neighbor, i.e. a vertex of degree 1, then performing a breadth-first-search starting with this individual to find the $\eta$ closest individuals in the network to this individual. Because of the spatial and computational limitations of Ripser we choose $600 \leq \eta \leq 3,000$ to ensure we can compute the persistence intervals of these sampled networks. In total we sampled from four different genealogical networks and four different social networks. These are the Advogat, LastFM Asia, Deezer HU and Deezer RO social networks and the genealogical networks 96--99 shown in Table \ref{Tab:2}, respectively. We sampled from each of these networks five times each to create a total of 20 sampled genealogical networks and 20 sampled social networks. The reason we begin our breadth-first search with a vertex of degree 1 is to ensure that our sampled networks have vertices both on the boundary and the interior of the original network we sampled to better mimic the structure of the original genealogical and social networks.

Apart from the (i) genealogical and social networks we consider and (ii) sampled versions of these networks, we also consider what we refer to as (iii) atypical genealogical networks. There are a number of genealogical networks that appear to be created with no attempt to represent all or even a fraction of the familial relationships. For example, the US Presidents network, cited as Atyp. Gen. Network 2 in Table \ref{Tab:2}, follows the shortest genealogical path between presidents leaving out extraneous relationships. We consider a number these \textit{atypical genealogical networks}, which form a contrast to the more standard genealogical networks we consider especially in terms of their peristent homology. A description of each of the (i) genealogical, social, (ii) sampled genealogical, sampled social, and (iii) atypical genealogical networks we consider is given at the end of the Appendix.

\begin{figure}[ht]
\begin{center}
    \begin{tabular}{cc}
    \includegraphics[width=.95\textwidth]{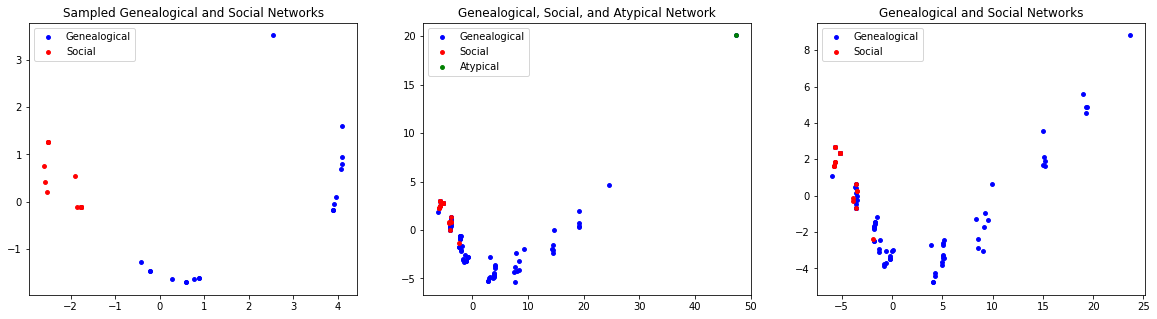}
\end{tabular}
\vspace{0.25cm}
\caption{PCA projections of the bottleneck distances between networks are shown. Left: The bottleneck distance between each of the twenty sampled genealogical and sampled social networks is shown. Center: The bottleneck distances are shown between the genealogical, social, and atypical genealogical networks we consider. Right: The bottleneck distances in the center panel are shown for only the genealogical and social networks we consider.}\label{fig:3}
\end{center}
\end{figure}

\section{Results}\label{sec:results}
Here we compare genealogical and other social networks using the (a) bottleneck distance and the (b) persistence curves defined in Section \ref{sec:4} (see Definitions \ref{def:bn} and \ref{def:percurv}, respectively). For those who have skipped Sections \ref{sec:3} and \ref{sec:4}, the bottleneck distance gives us a distance between two networks based on the differences in their persistent homology. Persistence curves give us a way of visualizing this difference but in greater detail (cf. Figure \ref{fig:persistence_curve}).

\subsection{Network Comparison using Bottleneck Distance}
\label{subsec:apply_bottleneck_distance}
Here we compute the bottleneck distance between every pair from the social and genealogical networks we consider. To visualize these results we use principal component analysis to identify the two components that account for the most variance and then plot this data in $\mathbb{R}^2$ (see Figure \ref{fig:3}).


From each part of Figure \ref{fig:3} we can see that genealogical networks are generally separated from social networks and form clusters that are easily distinguished. For the sampled networks (shown left), we can easily separate genealogical and social networks, and we can identify at least two distinct subclasses of genealogical networks. However, the bottleneck distance does an inferior job separating the non-sampled genealogical and social networks (shown center and right). The exception are the atypical genealogical networks, whose persistence intervals differ significantly enough from all of the other networks to be distinguishable as a third class of networks (shown center).


\subsection{Comparison of Genealogical and Social Networks using Persistence Curves}


\begin{figure}[ht]
\begin{center}
    \begin{tabular}{cc}
    \includegraphics[width=1.0\textwidth]{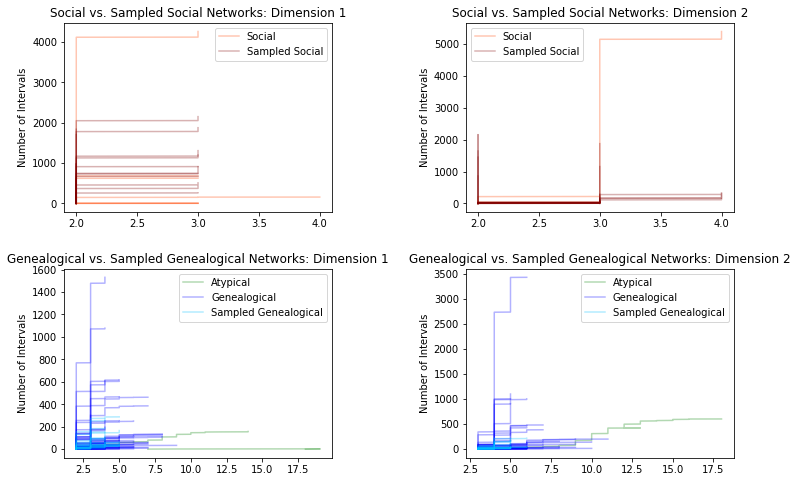}
\end{tabular}
\vspace{0.25cm}
\caption{Comparison of persistence curves for full networks vs sampled networks, grouped by dimension and type of network. Upper Row: Sampling social networks typically stretches the persistence curve in only one axis without affecting the other axis. Lower Row: Sampling genealogical networks typically shrink the persistence curve in both axes. Overall the average slope for social networks tends to increase when sampled, while genealogical networks experience a decrease in average slope.}\label{fig:curve_comp}
\end{center}
\end{figure}

Persistence curves give us a new alternative way of comparing networks. The advantage of using these curves compared to the bottleneck distance is that these curves give us a more detailed picture of how the number of persistence intervals varies from network to network. This allows us to better differentiate the structure of genealogical networks from social networks as well as observe the structure common to genealogical networks and those common to social networks, respectively.

In Figure \ref{fig:curve_comp} the persistence curves for the unsampled genealogical and unsampled social networks are shown in blue and red, respectively. The atypical genealogical networks are shown in green. The social networks have persistence curves that are quite vertical in both dimension 1 and dimension 2. For dimension 1, this indicates that most cycles in a social network are close to being trivial; either because they have a relatively small circumference or because they can be decomposed into a union of cycles with small circumferences. In particular, most of the social networks have a maximum death time of three (see Definition \ref{def:BD}), which corresponds to having a basis of cycles whose maximal circumference is at most nine. In other words, any cycle of circumference ten or more decomposes as the union of smaller cycles. For dimension 2, the steepness of the persistence curves indicate the presence of many distinct, yet similar, paths between certain pairs of vertices.

In contrast, the genealogical networks have persistence curves that have a much more horizontal profile indicating that most cycles are quite long and there are fewer `alternate paths' between pairs of vertices. In the extreme, the atypical genealogical networks are nearly flat in dimension 1, which reflects the fact that these atypical networks were intentionally constructed to have very few cycles. In dimension 2, the atypical networks show a similar slope to most of the typical genealogical networks, but the size of the alternative paths in these networks are much larger. This is likely due to the high number of individuals who were added only to link distant individuals, e.g. presidents. In a typical genealogical network, the additional relationships between such individuals would allow large cycles to decompose but in the atypical genealogical networks this in not the case.

\begin{figure}[ht]
\begin{center}
    \begin{tabular}{cc}
    \includegraphics[width=1.0\textwidth]{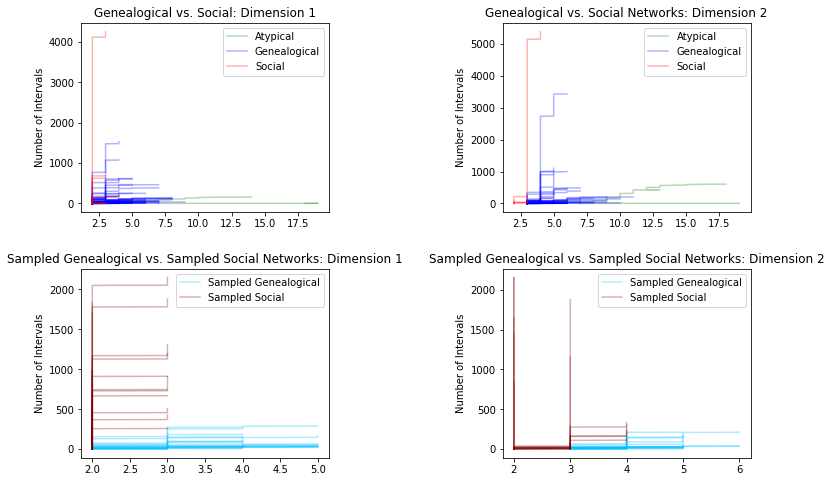}
\end{tabular}
\vspace{0.25cm}
\caption{Upper Row: Comparison of persistence curves for full networks by type. Lower Row: Comparison of persistence curves for sampled networks by type, excluding atypical genealogical networks. In each dimension, the average slope for genealogical networks is typically lower than the average slope for a social network. The atypical genealogical networks have the lowest average slope and much greater total length. The behavior for average slopes is more pronounced for sampled networks than for full networks.}\label{fig:curves_2}
\end{center}
\end{figure}


In Figure \ref{fig:curves_2}, we see the persistence curves for the sampled genealogical and sampled social networks shown in blue and red, respectively. The atypical genealogical networks are shown in green. Again the social networks have persistence curves that are quite vertical in both dimensions, although these curves are not as tall as in the case of unsampled social networks. This indicates that as a social network is sampled it retains a similar proportion of close-to-trivial cycles, but may lose many of the alternative paths between vertices that appear in dimension 2. By contrast, for genealogical networks the persistence curves indicate the complete loss of very large cycles in conjunction with a proportional loss of close-to-trivial cycles. In dimension 2, genealogical networks experience a more severe loss of alternative paths than the social networks. As a result, though sampling shrinks the scale of the persistence curves for social and genealogical networks, they remain visually distinct.


As in the bottleneck distance plots, genealogical and social networks appear to cluster together in that they have similar types of persistence curve. In fact, this is true whether or not the networks are sampled or unsampled. This suggests that even with incomplete data social network and genealogical networks have a distinguishable persistent homology, at least at the scales we consider.

It is worth mentioning that, while the bottleneck distance plots show us to an extent \textit{how} different genealogical and social networks are the persistence curves show us \textit{what} are differences are. The distance plots in Figure \ref{fig:3} do have the advantage of simplicity, however, and could presumably be used to more quickly identify differences in networks that are not as apparent as those we find between genealogical and social networks.

\subsection{Connections}
\label{subsec:connections}
It is also possible to use persistent homology to study properties of a network, such as the number of connected components, the typical size of cycles, or even ``missing links'' in the data. For genealogical and social networks, we can convert these mathematical concepts into more familiar ideas such as family groups or common ancestors. This also allows us to make conjectures about the persistent homology for such networks by converting standard assumptions about families or social networks into the language of persistence.

In dimension 0, the number of connected components determines the number of $\Pinterval{0}{\infty}$ intervals, and the total number of distinct vertices is the number of $\Pinterval{0}{\infty}$ intervals plus the number of $\Pinterval{0}{1}$ intervals. In the context of a genealogical network, each connected component represents a family group that is not related to the other family groups by any known connection. Thus, if a given family network is indeed a single ``family" of relatives, there should be exactly one $\Pinterval{0}{\infty}$ interval. In our Tikopia example we have eight $\Pinterval{0}{\infty}$ intervals each of which correspond to exactly one connected component of this genealogical network. (Note that Figure \ref{fig:1} (left) shows only the largest of these components). In this example, most of the the other `family groups' are actually individuals with no relation edges in the network.

In social networks, the connected components create what could be referred to as friend groups. Unlike genealogical networks, there are usually few restrictions on which edges form in a social network. As such, we do not have a conjecture about the number of $\Pinterval{0}{\infty}$ intervals in this setting in general. However, sampling any network as described in Section \ref{sec:5} will result in a new network with a single $\Pinterval{0}{\infty}$ interval.

\begin{figure}[ht]
    \begin{tabular}{c}
    \begin{overpic}[scale=0.185]{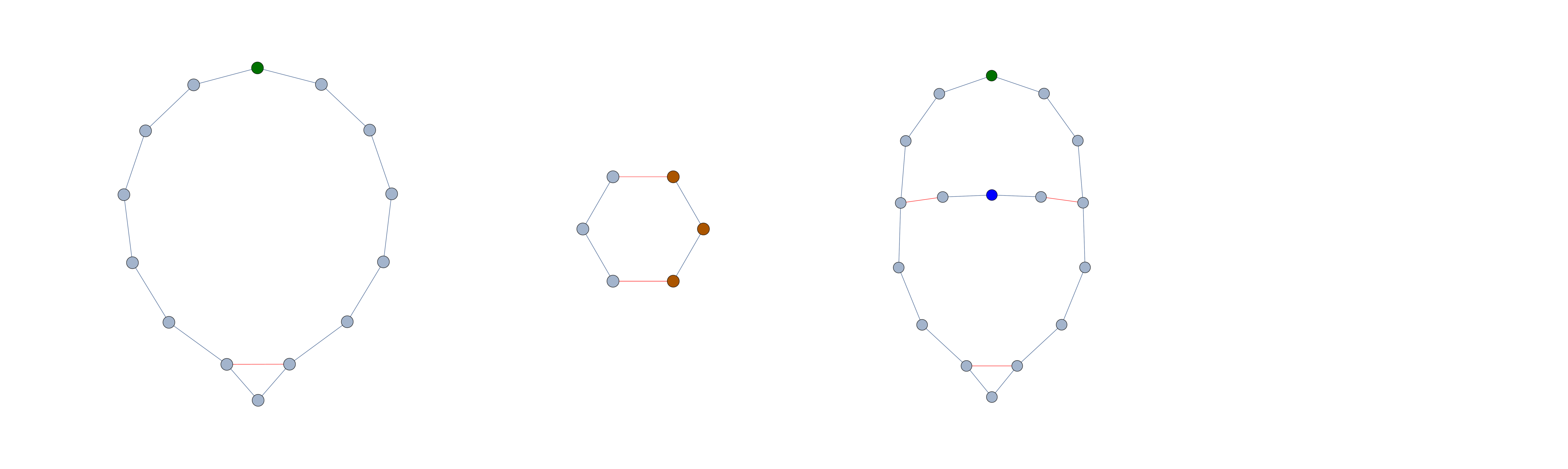}
    \put(6,0){\small{(a) Common Ancestor Cycle}}
    \put(35,0){\small{(b) \Horcycle}}
    \put(59,0){\small{(c) $\theta$-Cycle}}
    \end{overpic}
\end{tabular}
\caption{Left: A common ancestor cycle. The top most vertex is a common ancestor of the lowest vertex. The horizontal red line is a marriage, all other lines are parent-children edges. %
Center: A \horcycle, specifically the double cousin situation described in Section \ref{sec:2}. The left-most and right-most vertices are parents of their neighboring vertices. The two horizontal red lines are marriage edges. Right: A $\theta$-cycle formed by a common ancestor cycle with two overlapping hybrid cycles.}\label{fig:curves_1} 
\end{figure}

Moving to dimension 1, persistence intervals in this dimension describe the way that each connected component is internally structured. In sufficiently large genealogical networks, we will see three kinds of features that we call common ancestors, \horcycle s, and hybrid cycles. A \textit{common ancestor cycle} occurs when two descendants of an individual form a union or have a child together. We use the term \textit{\horcycle} to refer to situations where a cycle is formed through union edges and edges connecting two siblings. The final type of cycle of note, the \textit{hybrid cycles}, are those formed by any other combination of parent-child edges and union edges, which includes everything that is not a strict common ancestor or \horcycle. These three types of cycles are illustrated in Figure \ref{fig:curves_1}, where marriage edges are indicated by red edges and parent-child edges are indicated by blue edges. We show a common ancestor in Figure \ref{fig:curves_1}(a). Figure \ref{fig:curves_1}(b) is an example of a union cycle in which two siblings in one family form unions with two siblings in another, where only a single parent in each family is shown. In Figure \ref{fig:curves_1}(c) we give an example of a $\theta$-cycle, which is the union of a common ancestor cycle and two overlapping hybrid cycles. This example  comes from siblings of one family marrying cousins from another family. These cycles can be any length theoretically, but cultural norms affect the typical size and number of each type of cycle differently. Recording practices and incomplete data also limit whether these cycles appear in a given dataset. Thus having a description of these cycles together with an understanding of the culture may help identify errors in the recorded data. Conversely, understanding the distribution of cycles in high fidelity datasets can help identify the underlying cultural norms and help extrapolate where individuals are missing in incomplete data sets.

Since many cultures avoid marrying close relatives, common ancestor cycles tend to have a fairly large circumference. In the Tikopia network (see Figure \ref{fig:1}) we see persistence intervals with death values as high as 7 corresponding to cycles with a circumference of at least 21 individuals, which appear to be common ancestor cycles. This partially explains why persistence curves are so flat: there are relatively few minimal common ancestor cycles in a network, but they have very high persistence. More precisely, if the distance to union (the total number of individuals in a common ancestor cycle) is $n$, then the persistence of that cycle is $\lfloor n/3 \rfloor$. However, the \PHreps\ of persistent homology only include a basis for these cycles, instead of including every possible distinct cycle. In particular, a large common ancestor cycle will decompose into the union of two hybrid cycles if the hybrid cycles are each shorter than the common ancestor cycle, as shown in Figure \ref{fig:curves_1}(c). Persistent homology will reflect the size of the two smaller cycles instead of the larger common ancestor cycle. We note that it is possible to identify the actual cycles chosen for our basis, but the software we used does not provide that information and size of the networks prohibits us from identifying the cycles manually.

In social networks, we see that highly persistence cycles are quite rare. In order to have a cycle of persistence $3$, for instance, we need a loop with circumference 9 or higher with no shorter paths between any two vertices in the loops. It may be that phenomena like the small-world effect or, more colloquially, six-degrees of freedom limit the maximal persistence of social networks. We see this reflected in our example data sets with a maximum persistence of 3 for all but one of the social networks.







\section{Conclusion}\label{sec:conc}

In this paper, we explore the persistent homology structure of genealogical networks, motivated by the observation that family links tend to form in a fixed range of intermediate distances, which makes genealogical networks homologically distinct from most other social networks. We also introduce the notion of a persistence curve, which can be used to summarize and compare the persistent homology structure of any network. We also relate specific genealogical structures, such as the common ancestor cycle, to homology objects.

We find that, in the presence of incomplete data homology analysis is still genealogically useful. We note missing data due to recording practices and incomplete data (a ubiquitous feature of real genealogical networks), limits the kind of cycles that appear in a given dataset. Thus having a description of these cycles together with an understanding of the culture may help identify errors in the recorded data. Conversely, understanding the distribution of cycles in high fidelity datasets can help identify the underlying cultural norms and help extrapolate where individuals are missing in incomplete data sets.

There are several interesting directions in which this work could be expanded. For example, our work has made it clear that there is a real need to analyze the persistent homology of large networks, with at least tens of thousands of nodes, since family formation generally takes place at these scales. The Ripser library we relied on was not able to reach these scales. Additionally, we are very interested in creating random graph models which reflect the actual homology of human family networks---a first attempt at this by our group has been fairly successful at the scale of hundreds of nodes~\cite{rebeccathesis}. More broadly, there is a need to model the ground truth human family network. All the extant data sources represent biased, limited, and noisy subnetworks, while the true interest of the genealogical community is in the ground truth network. Tools for signal denoising, image inpainting, and graph extrapolation, for example, could be useful in this context. Finally, an important aspect of genealogical networks is the relationship between various supporting documents/metadata and the links that are discoverable through them. For example, one can consider optimal document collection strategies with a limited budget or document collection that is fair in terms of capturing minority information, which is often underrepresented.\\

\section{Declarations}

\subsection{Availability of data and materials}
\noindent Links to the datasets generated and/or analysed during the current study can be found in Table \ref{Tab:2}. Code to replicate and extend this work can be found at https://github.com/AbigailJ32/The-persistent-homology-of-genealogical-networks.

\subsection{Competing interests}
\noindent The authors declare that they have no competing interests.

\subsection{Funding}
\noindent ZB, BW, and AJ, were supported by a BYU CPMS CHIRP grant. ZB was additionally supported by NFS award \#2137511 and Army Research Office grant \#W911NF-18-1-0244, and the James S. McDonnell Foundation 21st Century Science Initiative—Complex Systems Scholar Award grant \#2200203. BW was additionally supported by the Simons Foundation grant \#714015. The views and conclusions contained in this document are those of the authors and should not be interpreted as representing the official policies, either expressed or implied, of the Army Research Office or the U.S. Government. The U.S. Government is authorized to reproduce and distribute reprints for Government purposes notwithstanding any copyright notation herein.

\subsection{Authors' contributions}
\noindent  Designed the experiments: ZB, NC, BW, RW. Performed the experiments: RF, RW. Wrote the paper: ZB, NC, TG, AJ, RS, BW, RW. All authors read and approved the final manuscript.

\subsection{Acknowledgements} \noindent We acknowledge helpful conversations with Joseph Price and the FamilySearch Engineering Research team. We also acknowledge Kolton Baldwin for helping to improve our code and simulations.

\section{Appendix}

Here we indicate both the genealogical and social networks used in our persistent homology computations (see Section \ref{sec:results}). We distinguish the datasets by network type: Friendship/Acquaintance, Social Media, Collaboration/Business, Disease Transmission, Information Sharing, Genealogical, and Atypical Genealogical networks.  We also provide the network name, number of vertices and edges in the network, and a citation where the network can be found. Also, a special thanks to Kolton Baldwin for help with numerical simulations on this paper.
{
\footnotesize
\begin{longtable}[c]{| c | c | c | l |}
\caption{Social and Genealogical Network Data Sets.\label{Tab:2}}\\
 \hline
 \multicolumn{4}{|c|}{\textbf{Network Data}}\\
 \hline
 Network Type $\&$ Name & Vertices & Edges & Citation\\
 \hline
 \endfirsthead

 \hline
 \multicolumn{4}{|c|}{}\\
 \hline
 Network Name $\&$ Type & Vertices & Edges & Citation\\
 \hline
 \endhead

 \hline
 \endfoot

 \hline
 \multicolumn{4}{| c |}{}\\
 \hline\hline
 \endlastfoot

 \hline
 \hline
 \multicolumn{4}{|c|}{\textbf{Social Networks}}\\
 \hline
    \textbf{Friendship $\&$} &&&\\
    \textbf{Aquaintance} &&&\\
\hline
    Dolphins& 62 & 159 & http://www-personal.umich.edu/~mejn/netdata/\\
    Zachary Karate Club&34&78&http://vlado.fmf.uni-lj.si/pub/networks/data/ucinet/ucidata.htm\#zachary\\ 
    Residence Hall & 217 & 2672 & http://konect.cc/networks/moreno\_oz/\\
    Highland Tribes & 16 & 58 & http://konect.cc/networks/ucidata-gama/\\
    Seventh Graders & 29 & 376 & http://konect.cc/networks/moreno\_seventh/\\
    Physicians & 241 & 1098 & http://konect.cc/networks/moreno\_innovation/ \\
    Highschool&70&366& http://konect.cc/networks/moreno\_highschool/\\
    Dutch College & 32 & 354 & http://konect.cc/networks/moreno\_vdb/\\
    Sampson's monastery & 25 & 322 & http://vladowiki.fmf.uni-lj.si/doku.php?id=pajek:data:esna3:sampson\\ 
    Adolescent health & 2539 & 12969 & http://konect.cc/networks/moreno\_health/\\
    Hamsterster friends & 2952 & 12534 & http://konect.cc/networks/petster-hamster-friend/\\ 
    Social Network 1 & 32 & 220 & http://vlado.fmf.uni-lj.si/pub/networks/doc/ECPR/assign.1/as1.net\\ 
    Social Network 2 & 32 & 191 & http://vlado.fmf.uni-lj.si/pub/networks/doc/ECPR/assign.1/as2.net\\ 
    Social Network 5 & 32 & 90 & http://vlado.fmf.uni-lj.si/pub/networks/doc/ECPR/assign.1/as5.net\\
    Social Network 7 & 32 & 61 & http://vlado.fmf.uni-lj.si/pub/networks/doc/ECPR/assign.1/as7.net\\
    Social Network 8 & 32 & 79 & http://vlado.fmf.uni-lj.si/pub/networks/doc/ECPR/assign.1/as8.net\\
    Social Network 9 & 32 & 58 & http://vlado.fmf.uni-lj.si/pub/networks/doc/ECPR/assign.1/as9.net\\
\hline
 \hline
    \textbf{Social Media} &&&\\
\hline
    Firm Hi-Tech & 33 & 124.5 & https://networkrepository.com/soc-firm-hi-tech.php\\
    Wiki-Vote & 889 & 2.9K & https://networkrepository.com/soc-wiki-Vote.php\\
    FB-PAGES-FOOD & 620 & 2.1K & https://networkrepository.com/fb-pages-food.php\\
    Advogato & 6541 & 51127 & http://konect.cc/networks/advogato/\\
    LastFM Asia & 7624 & 27806 & https://snap.stanford.edu/data/feather-lastfm-social.html\\
    Deezer HU & 47538 & 222887 & https://snap.stanford.edu/data/gemsec-Deezer.html\\
    Deezer RO & 41773 & 125826 & https://snap.stanford.edu/data/gemsec-Deezer.html\\

\hline
 \hline
    \textbf{Collaboration $\&$} &&&\\
    \textbf{Business} &&&\\
\hline
    Social Network 4 & 32 & 218 & http://vlado.fmf.uni-lj.si/pub/networks/doc/ECPR/assign.1/as4.net\\
    Social Network 6 & 32 & 103 & http://vlado.fmf.uni-lj.si/pub/networks/doc/ECPR/assign.1/as6.net\\
    Social Network 11 & 32 & 83 & http://vlado.fmf.uni-lj.si/pub/networks/doc/ECPR/assign.1/as11.net\\
    Social Network 12 & 32 & 65 & http://vlado.fmf.uni-lj.si/pub/networks/doc/ECPR/assign.1/as12.net\\

\hline
 \hline
    \textbf{Disease Transmission} &&&\\
\hline

    Taro Exchange & 22 & 78 & http://konect.cc/networks/moreno\_taro/\\

\hline
 \hline
    \textbf{Information Sharing} &&&\\
\hline
    Social Network 3 & 32 & 119 & http://vlado.fmf.uni-lj.si/pub/networks/doc/ECPR/assign.1/as3.net\\
    Social Network 10 & 32 & 80 & http://vlado.fmf.uni-lj.si/pub/networks/doc/ECPR/assign.1/as10.net\\

\hline
 \hline
 \multicolumn{4}{|c|}{\textbf{Genealogical Networks}}\\
 \hline
    Genealogical Network 1 & 310 & 322 & https://www.kinsources.net/kidarep/dataset-209-mowanjum-kalumburu.xhtml\\
    Genealogical Network 2 & 303 & 537 & https://www.kinsources.net/kidarep/dataset-2-mbuti-village-1957-af03.xhtml\\
    Genealogical Network 3 & 371 & 718 &   https://www.kinsources.net/kidarep/dataset-58-ojibwa-1930-nd07.xhtml\\
    Genealogical Network 4 & 795 & 1387 & https://www.kinsources.net/kidarep/dataset-150-achuar-pastaza.xhtml\\
    Genealogical Network 5 & 636 & 1151 &  https://www.kinsources.net/kidarep/dataset-92-chenchu-1940-as02.xhtml\\
    Genealogical Network 6 & 782 & 1366 &  https://www.kinsources.net/kidarep/dataset-28-trio-1960s.xhtml\\
    Genealogical Network 7 & 128 & 202 & https://www.kinsources.net/kidarep/dataset-23-shoshone-1880-nd11.xhtml\\
    Genealogical Network 8 & 439 & 626 & https://www.kinsources.net/kidarep/dataset-70-genesis.xhtml\\
    Genealogical Network 9 & 244 & 481 & https://www.kinsources.net/kidarep/dataset-66-waimiri-atroari.xhtml\\
    Genealogical Network 10 & 410 & 746 & https://www.kinsources.net/kidarep/dataset-240-kodiak.xhtml\\
    Genealogical Network 11 & 337 & 572 & https://www.kinsources.net/kidarep/dataset-51-wilcania.xhtml\\
    Genealogical Network 12 & 216 & 378 & https://www.kinsources.net/kidarep/dataset-22-ainu-1880-as01.xhtml\\
    Genealogical Network 13 & 77 & 134 & https://www.kinsources.net/kidarep/dataset-69-slavey-1911-nd12.xhtml\\
    Genealogical Network 14 & 815 & 1582 & https://www.kinsources.net/kidarep/dataset-7-pakaa-nova.xhtml\\
    Genealogical Network 15 &20&28&  https://www.kinsources.net/kidarep/dataset-38-wanindiljaugwa-1948-au06.xhtml\\
    Genealogical Network 16 &219&371& https://www.kinsources.net/kidarep/dataset-171-suya.xhtml\\
    Genealogical Network 17 &17&24& https://www.kinsources.net/kidarep/dataset-31-family.xhtml\\
    Genealogical Network 18 &168&221&  https://www.kinsources.net/kidarep/dataset-14-labrador-inuit-1776-nu02.xhtml\\
    Genealogical Network 19 &64&109& https://www.kinsources.net/kidarep/dataset-91-takamiut-1927-64-nu03.xhtml\\
    Genealogical Network 20 &1423&3211&  https://www.kinsources.net/kidarep/dataset-258-todas.xhtml\\
    Genealogical Network 21 &645&1097& https://www.kinsources.net/kidarep/dataset-65-igluligmiut-1961-nu07.xhtml\\
    Genealogical Network 22 &4463&8416& https://www.kinsources.net/kidarep/dataset-115-charlevoix.xhtml\\
    Genealogical Network 23 &48&86&  https://www.kinsources.net/kidarep/dataset-41-vedda-1905-as04.xhtml\\
    Genealogical Network 24 &104&172&
    https://www.kinsources.net/kidarep/dataset-71-igluligmiut-1960-61-nu08.xhtml\\
    Genealogical Network 25 &1263&2021&  https://www.kinsources.net/kidarep/dataset-223-samburu.xhtml\\
    Genealogical Network 26 &80&132&  https://www.kinsources.net/kidarep/dataset-10-apache-1932-nd01.xhtml\\
    Genealogical Network 27 &1269&2395&  https://www.kinsources.net/kidarep/dataset-24-ayd-nl-yoruk-2005.xhtml\\
    Genealogical Network 28 &299&532& https://www.kinsources.net/kidarep/dataset-13-tory.xhtml\\
    Genealogical Network 29 &19&30& https://www.kinsources.net/kidarep/dataset-21-ngatatjara-1966-au04.xhtml\\
    Genealogical Network 30 &399&592& https://www.kinsources.net/kidarep/dataset-204-dogon-konsogu-donyu.xhtml\\
    Genealogical Network 31 &377&712& https://www.kinsources.net/kidarep/dataset-49-alyawarra-1971-au01.xhtml\\
    Genealogical Network 32 &1263&2021&	 https://www.kinsources.net/kidarep/dataset-223-samburu.xhtml\\
    Genealogical Network 33 &118&192&	 https://www.kinsources.net/kidarep/dataset-39-eyak-1890.xhtml\\
    Genealogical Network 34 &98&161& https://www.kinsources.net/kidarep/dataset-75-nunamiut-1885-nu11.xhtml\\
    Genealogical Network 35 &479&830& https://www.kinsources.net/kidarep/dataset-19-ojibwa-1949-nd08.xhtml\\
    Genealogical Network 36 &1695&3206& https://www.kinsources.net/kidarep/dataset-103-tikuna-arara.xhtml\\
    Genealogical Network 37 &256&441& https://github.com/AbigailJ32/The-persistent-homology-of-genealogical-networks\\
    Genealogical Network 38 &798&1416&	 https://www.kinsources.net/kidarep/dataset-229-nucoorilma-tingha.xhtml\\
    Genealogical Network 39 &738&1212& https://www.kinsources.net/kidarep/dataset-32-yaraldi.xhtml\\
    Genealogical Network 40 &525&855&https://github.com/AbigailJ32/The-persistent-homology-of-genealogical-networks\\
    Genealogical Network 41 &619&1224&	 https://www.kinsources.net/kidarep/dataset-251-nunivak.xhtml\\
    Genealogical Network 42 &3008&6074& https://www.kinsources.net/kidarep/dataset-80-torshan.xhtml\\
    Genealogical Network 43 &278&464&	 https://www.kinsources.net/kidarep/dataset-62-dogrib-1911-25-59-nd04.xhtml\\
    Genealogical Network 44 &105&172&	 https://www.kinsources.net/kidarep/dataset-5-konkama-1931-44-51-eu02.xhtml\\
    Genealogical Network 45 &240&395& https://www.kinsources.net/kidarep/dataset-158-tikar.xhtml\\
    Genealogical Network 46 &4178&7351& https://www.kinsources.net/kidarep/dataset-45-obidos.xhtml\\
    Genealogical Network 47 &216&286&	 https://www.kinsources.net/kidarep/dataset-254-port-keats.xhtml\\
    Genealogical Network 48 &147&242&	 https://www.kinsources.net/kidarep/dataset-78-pul-eliya-1954-simpler-version.xhtml\\
    Genealogical Network 49 &277&516& https://www.kinsources.net/kidarep/dataset-213-sarmi.xhtml\\
    Genealogical Network 50 &330&622&	 https://www.kinsources.net/kidarep/dataset-73-parakana.xhtml\\
    Genealogical Network 51 &35&53& https://www.kinsources.net/kidarep/dataset-81-gundangborn-1948-au02.xhtml\\
    Genealogical Network 52 &48&76& https://www.kinsources.net/kidarep/dataset-84-hare-1956-nd05.xhtml\\
    Genealogical Network 53 &105&245& https://www.kinsources.net/kidarep/dataset-87-arara.xhtml\\
    Genealogical Network 54 &116&220& https://www.kinsources.net/kidarep/dataset-89-nunamiut-1960-nu13.xhtml\\
    Genealogical Network 55  &116&176&	 https://www.kinsources.net/kidarep/dataset-226-jie.xhtml\\
    Genealogical Network 56 &657&1166& 	 https://www.kinsources.net/kidarep/dataset-27-nyungar.xhtml\\
    Genealogical Network 57 &659&1288& https://www.kinsources.net/kidarep/dataset-3-anuta-1972.xhtmlj\\
    Genealogical Network 58 &112&182&  https://www.kinsources.net/kidarep/dataset-15-oodnadatta.xhtml\\
    Genealogical Network 59 &218&353&	 https://www.kinsources.net/kidarep/dataset-17-lainiovouma-1952-eu03.xhtml\\
    Genealogical Network 60 &90&119& https://www.kinsources.net/kidarep/dataset-12-miwuyt-1967-au03.xhtml\\
    Genealogical Network 61 &289&477&  https://www.kinsources.net/kidarep/dataset-9-konkama-1951-eu01.xhtml\\
    Genealogical Network 62 &1463&1969&  https://www.kinsources.net/kidarep/dataset-306-nobles-ile-de-france-1000-1440.xhtml\\
    Genealogical Network 63 &4109&6517&  https://www.kinsources.net/kidarep/dataset-287-duu-rea.xhtml\\
    Genealogical Network 64 &29&48&  https://www.kinsources.net/kidarep/dataset-46-hatfields-and-mccoys.xhtml\\
    Genealogical Network 65 &40&59&  https://www.kinsources.net/kidarep/dataset-33-angmagsalik-1884-nu01.xhtml\\
    Genealogical Network 66 &294&441&  https://www.kinsources.net/kidarep/dataset-18-tikopia-1930.xhtml\\
    Genealogical Network 67 &502&786&  https://www.kinsources.net/kidarep/dataset-34-netsilik-1922-nu09.xhtml\\
    Genealogical Network 68 &83&126&  https://www.kinsources.net/kidarep/dataset-8-semang-1924-50-as03.xhtml\\
    Genealogical Network 69 &95&157&  https://www.kinsources.net/kidarep/dataset-4-shoshone-1860-nd10.xhtml\\
    Genealogical Network 70 &2588&5651&  https://www.kinsources.net/kidarep/dataset-61-kelkummer.xhtml\\
    Genealogical Network 71 &88&144&  https://www.kinsources.net/kidarep/dataset-77-apache-1935-nd02.xhtml\\
    Genealogical Network 72 &1513&2217&  https://www.kinsources.net/kidarep/dataset-90-omaha-1880.xhtml\\
    Genealogical Network 73 &3014&5454&  https://www.kinsources.net/kidarep/dataset-128-ammonni.xhtml\\
    Genealogical Network 74 &139&201&  https://www.kinsources.net/kidarep/dataset-79-paiute-1880-nd09.xhtml\\
    Genealogical Network 75 &5016&10719&  https://www.kinsources.net/kidarep/dataset-249-baruya.xhtml\\
    Genealogical Network 76 &125&202&  https://www.kinsources.net/kidarep/dataset-242-tlingit.xhtml\\
    Genealogical Network 77 &272&445&  https://www.kinsources.net/kidarep/dataset-36-copper-1922-nu10.xhtml\\
    Genealogical Network 78 &378&609&  https://www.kinsources.net/kidarep/dataset-52-apache-1936-nd03.xhtml\\
    Genealogical Network 79 &926&1951&  https://www.kinsources.net/kidarep/dataset-68-surui.xhtml\\
    Genealogical Network 80 &706&1177&  https://www.kinsources.net/kidarep/dataset-60-mbuti-forest-1957-af02.xhtml\\
    Genealogical Network 81 &435&672&  https://www.kinsources.net/kidarep/dataset-64-melombo.xhtml\\
    Genealogical Network 82 &128&114&  https://www.kinsources.net/kidarep/dataset-164-kaingang.xhtml\\
    Genealogical Network 83 &169&275&  https://www.kinsources.net/kidarep/dataset-11-top-of-the-mountain.xhtml\\
    Genealogical Network 84 &178&274&  https://www.kinsources.net/kidarep/dataset-37-igluligmiut-1921-nu05.xhtml\\
    Genealogical Network 85  &87&111&  https://www.kinsources.net/kidarep/dataset-216-tiwi.xhtml\\
    Genealogical Network 86 &2049&4159&  https://www.kinsources.net/kidarep/dataset-35-chuukese-1947-1940.xhtml\\
    Genealogical Network 87 &868&980&   https://www.kinsources.net/kidarep/dataset-20-saudi-royal-genealogy.xhtml\\
    Genealogical Network 88 &2821&5079&  https://www.kinsources.net/kidarep/dataset-30-manus-1929.xhtml\\
    Genealogical Network 89 &454&980&  https://www.kinsources.net/kidarep/dataset-74-arawete.xhtml\\
    Genealogical Network 90 &304&472&  https://www.kinsources.net/kidarep/dataset-42-nunamiut-tareumiut-1900-nu12.xhtml\\
    Genealogical Network 91 &367&671&  https://www.kinsources.net/kidarep/dataset-48-wanindiljaugwa-1941-au05.xhtml\\
    Genealogical Network 92 &3151&4289&  https://www.kinsources.net/kidarep/dataset-54-feistritz-am-gael-1990.xhtml\\
    Genealogical Network 93 &2975&5107&  https://www.kinsources.net/kidarep/dataset-159-cocama-cocamilla.xhtml\\
    Genealogical Network 94 &585&1249&  https://www.kinsources.net/kidarep/dataset-44-torres-strait.xhtml\\
    Genealogical Network 95 &334&530&  https://www.kinsources.net/kidarep/dataset-6-igluligmiut-1949-nu06.xhtml\\
    Genealogical Network 96 &9595&14988&  https://www.kinsources.net/kidarep/dataset-93-sainte-catherine.xhtml\\
    Genealogical Network 97 &28586&51446&  https://www.kinsources.net/kidarep/dataset-76-san-marino.xhtml \\
    Genealogical Network 98 &18645&32439&	 https://www.kinsources.net/kidarep/dataset-307-bwa-slam-biogsurvey.xhtml \\
    Genealogical Network 99 &8809&15643&  https://www.kinsources.net/kidarep/dataset-194-kel-owey.xhtml  \\
\hline
 \hline
    \textbf{Atypical Genealogical} &&&\\
    \textbf{Networks} &&&\\
\hline
    Atyp. Gen. Network 1 & 429 & 705 & (created using FamilySearch.org)\\ 
    Atyp. Gen. Network 2 & 2477 & 4015 &  https://www.kinsources.net/kidarep/dataset-56-us-presidents.xhtml

 \end{longtable}
}


\end{document}